\documentclass{ws-ijmpb}
\usepackage{graphicx}

\begin{document}

%\markboth{Authors' Names}
%{}
\markboth{B. Wolf, A. Honecker, W. Hofstetter, U. Tutsch, M. Lang}
{Cooling through quantum criticality \& many-body effects in
condensed matter and cold gases}

\catchline{}{}{}{}{}

%
%%%%%%%%%%%%%%%%%%%%%%%%%%%%%%%%%%%%%%%%%%%%%%%%%%%%%%%%%%%%%%%%%%%%

\title{Cooling through quantum criticality and many-body effects in
condensed matter and cold gases
%\\
%USING \LaTeX\footnote
%{For the
%title, try not to use more than 3 lines.
%Typeset the title in 10~pt Times Roman, uppercase and boldface.}
  }

\author{Bernd Wolf}

\address{Institute of Physics, Goethe Universit\"at, SFB/TR 49\\
Max-von-Laue Strasse 1, 60438 Frankfurt, Germany\\
wolf@physik.uni-frankfurt.de}

\author{Andreas Honecker\footnote{Address after September 1, 2014:
Laboratoire de Physique Th\'eorique et Mod\'elisation,
Universit\'e de Cergy-Pontoise,
2 avenue A.\ Chauvin,
95302 Cergy-Pontoise cedex, France.}}

\address{Institute for Theoretical Physics,
Georg-August-Universit\"at G\"ottingen \\
Friedrich-Hund-Platz 1, 37077 G\"ottingen, Germany \\
honecker@theorie.physik.uni-goettingen.de}

\author{Walter Hofstetter}
\address{Institute for Theoretical Physics, Goethe Universit\"at, SFB/TR 49\\
Max-von-Laue Strasse 1, 60438 Frankfurt, Germany%\footnote{State completely without abbreviations, the
%affiliation and mailing address, including country. Typeset in
%8~pt Times Italic.}
\\
hofstett@physik.uni-frankfurt.de%\footnote{Typeset author e-mail
%address in single line.}
}

\author{Ulrich Tutsch}

\address{Institute of Physics, Goethe Universit\"at, SFB/TR 49\\
Max-von-Laue Strasse 1, 60438 Frankfurt, Germany\\
tutsch@physik.uni-frankfurt.de}

\author{Michael Lang%\footnote{Typeset names in
%8~pt Times Roman, uppercase. Use the footnote to indicate the
%present or permanent address of the author.}
}

\address{Institute of Physics, Goethe Universit\"at, SFB/TR 49\\
Max-von-Laue Strasse 1, 60438 Frankfurt, Germany%\footnote{State completely without abbreviations, the
%affiliation and mailing address, including country. Typeset in
%8~pt Times Italic.}
\\
Michael.Lang@physik.uni-frankfurt.de%\footnote{Typeset author e-mail
%address in single line.}
}

\maketitle

\begin{history}
\received{29 August 2014}
%\revised{Day Month Year}
\accepted{7 September 2014}
Published 10 October 2014\par
%\comby{(xxxxxxxxxx)}
\end{history}

\begin{abstract}
%The abstract should summarize the context, content
%and conclusions of the paper in less than 200 words. It should
%not contain any references or displayed equations. Typeset the
%abstract in 8 pt Times Roman with baselineskip of 10~pt, making
%an indentation of 1.5 pica on the left and right margins.
This article reviews some recent developments for new cooling
technologies in the fields of condensed matter physics and cold
gases, both from an experimental and theoretical point of view. The
main idea is to make use of distinct many-body interactions of the
system to be cooled which can be some cooling stage or the material
of interest itself, as is the case in cold gases. For condensed
matter systems, we discuss magnetic cooling schemes based on a large
magnetocaloric effect as a result of a nearby quantum phase
transition and consider effects of geometrical frustration. For
ultracold gases, we review many-body cooling techniques, such as
spin-gradient and Pomeranchuk cooling, which can be applied in the
presence of an optical lattice. We compare the cooling performance
of these new techniques with that of conventional approaches and
discuss state-of-the-art applications.
\end{abstract}


%Contributions to {\it International Journal of Modern Physics B}
%will be reproduced by using the author's submitted typeset manuscript.
%It is therefore essential that the manuscript be in its final form, and
%of good appearance. The typeset manuscript should be submitted to the
%publisher in PDF format as well as in its \LaTeX\ format.

%Contributions are to be in English. Authors are encouraged to
%have their contribution checked for grammar. American spelling
%should be used. Abbreviations are allowed but should be spelt
%out in full when first used. Integers ten and below are to be
%spelt out. Italicize foreign language phrases (e.g.~Latin,
%French).

\section{Introduction}

The properties of matter at low temperatures continue to be of high
interest. Phenomena such as novel types of
superconductivity/superfluidity, quantum phase transitions or
different types of topological order fascinate a growing community
of researchers. Advancements in these different areas rely on
suitable coolants. Since standard cooling technologies, used in
experiments on condensed matter systems, are based on $^3$He which
is difficult in handling and of limited availability, alternative
cooling technologies are required. Likewise, with the advent of
ultracold gases as tunable quantum simulators for the
above-mentioned many-body effects, there is an increasing demand for
efficient cooling technologies in this rapidly developing field.

For experiments on \textbf{condensed matter systems}, P. Debye
\cite{Debye1926} and W. F. Giauque \cite{Giauque1927} independently
suggested in 1926 to use the magnetocaloric effect (MCE) of
paramagnetic materials to reach temperatures significantly below
1~K. This effect, which describes temperature changes of a magnetic
material in response to an adiabatic change of the magnetic field,
forms the basis of magnetic refrigeration with the main area of
applications focussing on cryogenic
temperatures.\cite{Lounasmaa1974,Tishin2003,Gschneidner2005} A
highly topical area of research has been triggered by the
observation of a giant MCE around room temperature.
\cite{Gschneidner1997} This discovery has stimulated additional work
\cite{Brueck2005,Tegus2002} indicating the potential of the MCE for
an environment-friendly room-temperature refrigeration.

A large MCE, and with it a large cooling effect, can be expected for
materials where the entropy $S$ strongly changes with the magnetic
field $B$. This is the case for paramagnetic salts, the materials of
choice for low-temperature refrigeration. \cite{Lounasmaa1974} These
systems are characterized by a three-dimensional (3D) array of
spin-carrying centres, with a sufficiently large inter-site distance
so that residual magnetic interactions are weak. Besides their large
$\partial S$/$\partial B$ values, paramagnets excel by their ease of
operation as compared to $^{3}$He-$^{4}$He dilution refrigerators,
the standard cooling technology for reaching sub-Kelvin
temperatures. In addition, paramagnets can be operated under
microgravity conditions, where standard dilution refrigerators fail,
and are thus vital in present space applications.
\cite{Shirron2007,Hagemann1999}

An extension of the concept of paramagnets includes molecular
nanomagnets, where the magnetic centres consist of extended,
magnetically isolated molecules. Recently it has been recognized
that these systems can exhibit a large MCE at low temperatures which
makes them attractive for magnetic cooling.
\cite{Spichkin2001,Sessoli2012} High-spin single molecules with a
large spin value (superparamagnets), having a small or vanishing
anisotropy, are particularly favourable. The reason for that is the
large corresponding magnetic entropy of these molecules which can be
easily polarized in a magnetic field. \cite{Evangelisti2006} Some of
the superparamagnets are also geometrically frustrated and therefore
exhibit large changes of their magnetization below the saturation
field, in favour of a large MCE. \cite{Schnack2007,Honecker2009}

An alternative to the above paramagnetic systems is provided by
geometrically frustrated magnets,
\cite{Ramirez1994,Schiffer1996,Greedan2001,Moessner2001,Lacroix2011}
characterized by considerable inter-site magnetic interactions which
compete with each other. Starting with the recognition that the
magnetocaloric substance Gd$_3$Ga$_5$O$_{12}$
\cite{Fisher1973,Hornung1974,Brodale1975,Barclay1982} owes its large
MCE to geometric frustration,
%\cite{Zhitomirsky2003a}
experimental and theoretical studies
% over the last two decades
have established these systems as powerful coolants for temperatures
in the Kelvin range with a minimum accessible temperature around
0.5~K. Despite the interaction between neighbouring spins, strongly
frustrated magnets remain in a so-called disordered cooperative
paramagnetic state with finite entropy at temperatures well below
the materials' Curie-Weiss temperatures. \cite{Villain1971} In this
temperature range and under suitable conditions the cooling rate
($\partial T/\partial B)_{S}$ for classical spin systems can exceed
the values obtained for paramagnets by more than one order of
magnitude. \cite{Zhitomirsky2003} The enhanced cooling rate of
geometrically frustrated magnets in the cooperative paramagnetic
state is due to the magnetic entropy which results from a
macroscopic number of local modes that at the classical level remain
gapless up to the saturation field. \cite{Sosin2005} In contrast,
for non- or weakly-frustrated antiferromagnets, the condensation of
only one or a few modes leads to long-range order described by a
certain wave vector. \cite{Zhitomirsky2003} Remarkably, due to exact
``localized magnon'' states,
\cite{Schnack2001,Schulenburg2002,Zhitomirsky2004,ZhiTsu04,ZhiTsu05,Derzhko2007}
the classical degeneracy and thus the good magnetocaloric properties
survive quantum fluctuations at the saturation field.

The  well-established ``classical'' cooling systems mentioned
further above have been extensively discussed in the literature and
there are numerous excellent review articles available, see, e.g.,
Refs. \cite{Brueck2005,Ambler1955,Pecharsky1999}. In the present paper we focus on an alternative Ansatz
taken up more recently. Here a strongly field-dependent entropy, and
with it a large MCE, is the result of distinct many-body
interactions. These are materials close to a quantum-critical point
(QCP) -- a zero-temperature phase transition -- separating an
ordered from a quantum-disordered state. For these materials the
low-temperature properties are governed by strong quantum many-body
effects which give rise to a peculiar entropy landscape. The
transition to saturation in highly frustrated magnets mentioned
before can be considered as one example for this category. These new
coolants may provide a useful alternative for experiments in the
sub-Kelvin temperature range where they have the potential to
replace established cooling technologies based on $^{3}$He.

In \textbf{ultracold gases}, a major current challenge is reaching the low temperatures (entropies) required for observing solid-state type ordered
states, for example quantum magnetism induced by superexchange, or $d$-wave superfluidity,
which would allow highly tunable quantum simulations of model systems
such as the fermionic Hubbard model\cite{Hofstetter2002,Bloch2008} or Heisenberg spin models.
While Mott-insulating phases\cite{Joerdens2008,Schneider2008} short-range magnetic correlations\cite{Greif2013}
and tunable magnetic exchange couplings\cite{Trotzky2008}
have already been realized both in bosonic and fermionic multiflavor gases, true long-range magnetic order in optical lattices has so far
only been observed in a realization of the quantum XY model via a tilted bosonic Mott-insulator, where the effective magnetic exchange
is linear in the hopping amplitude instead of quadratic\cite{Simon2011}. Current estimates of entropies per particle in a fermionic $^{40}$K Mott insulator yield $S/N \approx k_B \ln 2$, which is roughly twice the critical entropy for
N\'eel ordering\cite{Joerdens2010}.

In order to reach these low temperatures (entropies),
new cooling techniques in the presence of an optical lattice need to be implemented.
In contrast to solid-state materials,  ensembles of ultracold gases in optical lattices are to a very good approximation
closed quantum systems, where the total entropy remains constant if the time evolution,
e.g., during a lattice ramp, is adiabatic. The quantum phases under investigation, such as Mott insulator or magnetic states, are therefore uniquely characterized by their entropies per particle, while the corresponding temperatures depend on the strength of the
optical lattice potential and on the atomic scattering properties, and are
non-universal. In the following discussion we will, for this reason, mostly focus on entropy
instead of temperature.

The goal is then to either prepare a sufficiently low-entropy initial state, which upon (approximately) adiabatic ramp-up of the lattice
becomes magnetically ordered, or to separate entropy-rich from low-entropy regions of the sample, e.g., by shaping the optical potential or by magnetic gradients.
Both approaches will be discussed in section \ref{sec:cold_gases}.

The paper is organized as follows. After describing the basic
thermodynamic relations for magnetic cooling in section \ref{sec:2},
we discuss in section \ref{sec:3} different concepts by comparing
the conventional approach for magnetic cooling with the novel Ansatz
of cooling through many-body effects. Section \ref{sec:Qcrit} deals
with the magnetic cooling of real systems before typical
applications are discussed in section \ref{sec:appl}.

\section{Basic relations}

\label{sec:2}

We start by reviewing some basic properties of the magnetocaloric effect. For simplicity, we focus
on a purely magnetic system that can be described by a free energy $F(T,B,N)$
where $T$ is the temperature, $B$ the external magnetic field and $N$ the number
of spins (particles). In canonical thermodynamics the free energy is expressed
via the partition function $Z$ as $F=-k_B\,T\,\ln Z$. This is of course a very useful
relation in order to compute the thermodynamic quantities for a given model, but in this
section we rather want to recall that some useful results can already be derived on general
grounds.

The physical quantities of interest
are obtained by appropriate partial derivatives of
the free energy, in particular the \emph{entropy}
\begin{equation}
S = - \frac{\partial\, F}{\partial T} \, ,
\label{eq:defS}
\end{equation}
the {\em magnetization}
\begin{equation}
M = - \frac{\partial\, F}{\partial B} \, ,
\label{eq:defM}
\end{equation}
and the {\em specific heat}
\begin{equation}
C = T \,\frac{\partial\, S}{\partial T} \, .
\label{eq:defC}
\end{equation}
This last relation is already useful for analysing experiments since it allows one to
reconstruct the experimentally inacessible entropy from the experimentally
accessible specific heat \cite{Tishin2003}
in the form $\Delta S = \int \frac{C}{T} \, {\rm d}T$.

Now let us consider a change in an external magnetic field $B$. In general, the
system will respond with a change of temperature $T$. The \emph{magnetocaloric
effect} is defined as the corresponding derivative ${\partial T}/{\partial B}$.
If we consider perfectly adiabatic conditions, i.e., constant entropy $S$,
we can capture this effect by introducing a \emph{magnetic Gr\"uneisen
parameter}
\begin{equation}
\Gamma_{B} := \frac{1}{T} \, \left(\frac{\partial \,T}{\partial B}\right)_{S} \, .
\label{GammaB1}
\end{equation}
By virtue of Eq.~(\ref{eq:defS}), the entropy is also a function of $T$, $B$,
and $N$. Keeping $N$ fixed, and using
Eq.~(\ref{eq:defC}) we have
\begin{equation}
{\rm d}S = \frac{\partial \,S}{\partial T}\,{\rm d}T
         + \frac{\partial \,S}{\partial B}\,{\rm d}B
= \frac{C}{T}\,{\rm d}T
         + \frac{\partial \,S}{\partial B}\,{\rm d}B \, .
\label{eq:dS}
\end{equation}
Under adiabatic conditions we have ${\rm d}S = 0$, thus
\begin{equation}
0 = \frac{C}{T} \,  \left(\frac{\partial \,T}{\partial B}\right)_{S}
 +  \left(\frac{\partial \,S}{\partial B}\right)_{T} \, ,
\label{eq:dSdB}
\end{equation}
and hence with Eq.~(\ref{GammaB1})
\begin{equation}
\Gamma_{B} = -\frac{1}{C} \, \left(\frac{\partial \,S}{\partial B}\right)_{T}
= -\frac{\left({\partial \,S}/{\partial B}\right)_{T}}{
         T\,\left({\partial \,S}/{\partial T}\right)_{B}}
\, . \label{GammaB2}
\end{equation}
The analogy of the last form to a similar expression for the
classical, \textit{thermal Gr\"uneisen parameter} used in describing
the thermal expansion, motivates the name as well as the inclusion
of the factor $1/T$ into the definition (\ref{GammaB1}).
\cite{Zhu2003,Garst2005}

Finally, the derivative of the entropy with respect to magnetic field
can be eliminated using the definitions Eqs.~(\ref{eq:defS}) and
(\ref{eq:defM}) and one finds
\begin{equation}
\Gamma_{B} = -\frac{1}{C} \, \frac{\partial \,M}{\partial T}
\, . \label{GammaB3}
\end{equation}
This relation has been widely used experimentally
%\cite{Tishin2003}
since it relates the magnetocaloric effect to the specific heat and
the variation of magnetization with temperature at a given
temperature, i.e., quantities that are straightforward to measure
without the need to ensure adiabatic conditions.

As a first simple application let us consider an \emph{ideal paramagnet}
that we may characterize by an entropy that is a function of only the ratio
$B/T$ but not of both parameters individually, i.e., $S=S(B/T,N)$. It follows that
$C= -B\,\,\frac{\partial \,S}{\partial B}$ and thus with
Eq.~(\ref{GammaB2})
\begin{equation}
\Gamma_{B} = \frac{1}{B}
 \label{GammaBidealPara}
\end{equation}
for an ideal paramagnet. Actually, given that $S=S(B/T,N)$, the temperature $T$ has to vary
linearly with magnetic field $B$ in order to keep the entropy constant.
Thus, for an ideal paramagnet ${\partial T}/{\partial B}$ has to be constant
along an adiabatic demagnetization curve which is a straight line going to zero for $B \to 0$.
The slope of the line through $(0,0)$ and $(T,B)$ is
$\frac{T}{B}$, i.e.,
$\left(\frac{\partial \,T}{\partial B}\right)_S = \frac{T}{B}$. Upon division by
$T$, one recovers Eq.~(\ref{GammaBidealPara}) from these basic
considerations.

The behaviour of the \textit{magnetic Gr\"{u}neisen parameter} in
the vicinity of a {\it quantum critical point} -- a zero-temperature
phase transition -- is a bit more involved, but some conclusions
can already be reached with the tools above and simple scaling
assumptions. \cite{Zhu2003,Garst2005} Let us consider a
field-induced quantum phase transition at $B=B_c$, i.e., a
transition between two distinct phases that meet at $B=B_c$ for
$T=0$. Introducing the scaling parameter
\begin{equation}
r = \frac{B-B_c}{B_0}
 \label{eq:defR}
\end{equation}
with some field scale $B_0$, the singular part of the free energy density close to a
second-order quantum phase transition can be cast in the scaling form \cite{Garst2005}
\begin{equation}
f(B,T) = a^{-(d+z)} \, g(r\,a^{1/\nu},T\,a^z)
 \label{eq:fScale1}
\end{equation}
with $d$ the spatial dimension, $z$ the dynamical critical exponent, $\nu$ the correlation length
critical exponent, and $g$ a scaling function. Firstly, we can choose the arbitray scale parameter
$a=r^{-\nu}$ in order to find
\begin{equation}
f(B,T) = r^{(d+z)\,\nu} \, g(1,T\,r^{-z\,\nu})
=: r^{(d+z)\,\nu} \, \tilde{g}(T\,r^{-z\,\nu}) \, ,
 \label{eq:fScale2}
\end{equation}
where $\tilde{g}$ is a new one-parameter scaling function. If we are interested in
the low-temperature limit $T \to 0$, $r \ne 0$, we can
set $x=T\,r^{-z\,\nu}$ and expand \cite{Zhu2003}
\begin{equation}
\tilde{g}(x) = \tilde{g}(0) + C\,x^{y_0+1} + \ldots \, ,
 \label{eq:gExpand}
\end{equation}
where $y_0$ is another exponent whose value may depend on the side of the phase transition,
i.e., the value of $y_0$ may be different for $r<0$ and $r>0$.

Now it remains to analyse the derivatives of the free energy using
the scaling form Eq.~(\ref{eq:fScale2}) and the expansion (\ref{eq:gExpand}). \cite{Zhu2003,Garst2005}
First, one observes that the entropy density $S/V = -\partial f/\partial T$ scales as
$S/V \propto T^{y_0}$, i.e., $y_0 > 0$ is required in order to satisfy the third law of
thermodynamics. Finally, considering second derivatives and using Eq.~(\ref{GammaB2})
one finds the scaling form of the magnetic Gr\"uneisen parameter
\begin{equation}
\Gamma_{B} = -\frac{\left({\partial \,S}/{\partial B}\right)_{T}}{
         T\,\left({\partial \,S}/{\partial T}\right)_{B}}
= -G_r \, \frac{1}{B-B_c}
 \label{eq:GammaScale}
\end{equation}
with the prefactor given by
\begin{equation}
G_r = \frac{(d-y_0\,z)\,\nu}{y_0} \, .
 \label{eq:Gr}
\end{equation}
Remarkably, all unknown quantities cancel out and one is left with a
numerical prefactor that is a combination of {\em universal}
exponents only. \cite{Zhu2003,Garst2005}

We would like to note that the general low-temperature form Eq.~(\ref{eq:GammaScale})
holds not only at quantum phase transitions, but also for the ideal paramagnet
(compare Eq.~(\ref{GammaBidealPara})) as well as certain highly degenerate critical
points related to localized
magnons \cite{Schulenburg2002,Zhitomirsky2004,ZhiTsu04,ZhiTsu05,Derzhko2007}
where one can apply similar scaling arguments as above.

\section{Concepts for magnetic cooling}

\label{sec:3}

\begin{figure}[bt]
%\centerline{\psfig{file=Fig1_02062014.eps,width=3.65in}}
\centerline{\includegraphics[width=3.65in]{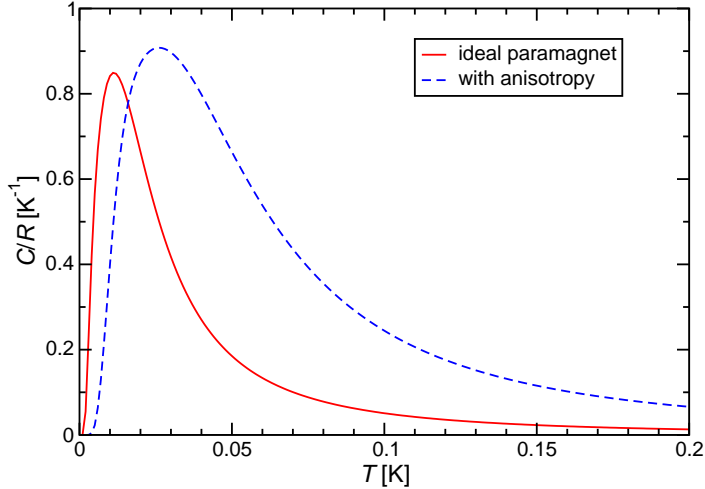}}
%\vspace*{8pt}
\caption{Specific heat of an ideal and a realistic spin-5/2 paramagnet
in a magnetic field $B = 10$~mT.
For the ideal paramagnet, the energy function is $E(S^z) = -g\,\mu_B\,B\,S^z$;
for the other case we use a model for the ferric ammonium alum with
 $E(S^z) = D_2\,(S^z)^2 + D_4\,(S^z)^4-g\,\mu_B\,B\,S^z$ amd
 $D_2/k_B = -85$~mK, $D_4/k_B =12$~mK \protect\cite{Kimura1967}. The spectroscopic $g$ factor is
 2 in both cases.
}
\label{fig:Cpara}
\end{figure}

As described in section \ref{sec:2}, the process of magnetic cooling is
closely related to the magnetic entropy $S_{\rm mag}$ of the coolant and
the way $S_{\rm mag}$ varies with magnetic field and temperature.
Generally, the precondition for an effective magnetic cooling is a
large absolute value of $S_{\rm mag}$ in the temperature range of
interest together with a large $\partial S/\partial B$. The
concept of magnetic cooling can be visualized particularly clearly
for paramagnets, whose total molar entropy is given by
\begin{equation}
S_{\rm mag} = R \cdot \ln(2s + 1)\, , \label{Smag}
\end{equation}
with $R$ the gas constant and $s$ the spin value. Due to weak magnetic interactions between
the ions, inevitable for any real material, the degeneracy is (at
least partly) lifted. As a consequence real paramagnets exhibit,
depending on their spin state, a Schottky-like anomaly, $C_{\rm mag}^{\rm Sch}$, in their
magnetic specific heat. The maximum in
$C_{\rm mag}^{\rm Sch}$ is located at a temperature $T_{\rm max}$ which
corresponds to about half of the zero-field splitting $T_{\rm max} \sim$
0.5 $\Delta/k_B$. Due to the Zeeman-effect, the level splitting
further increases with increasing field and the Schottky-like
anomaly is shifted to higher temperatures. In other
words, the corresponding variation of $S_{\rm mag}$ with $T$ and $B$ for a
paramagnetic material is determined by the zero-field splitting of
the magnetic energy levels and can easily be varied with an external
magnetic field.
% Since for non-interacting spins the entropy depends
%only on the ratio of magnetic energy ($\propto B)$ to thermal energy
%($\propto T$), i.e., $S$ = $S$($B/T$), a paramagnet exhibits a
%linear variation of temperature with magnetic field and a constant
%$\partial S$/$\partial B$ (wie kann man das erkennen?) which
%directly leads to a constant cooling rate.
This is illustrated in Fig.~\ref{fig:Cpara} with the specific heat
of an ideal paramagnet and a more realistic spin-5/2 model
paramagnet (with a small single-ion anisotropy) in an external
magnetic field of $B=10$~mT. On the one hand, the isentropes for the
ideal paramagnet would go to $T=0$ as $B \to 0$ while the lowest
accessible temperature in a real paramagnet is determined by the
zero-field gap. On the other hand, we infer from
Fig.~\ref{fig:Cpara} that the ability of an ideal paramagnet in a
magnetic field of 10~mT to exchange heat with a cooling load would
be limited to a narrow temperature window around 10~mK, while that
of a more realistic paramagnet extends to higher temperatures.

\begin{figure}[bt]
%\centerline{\psfig{file=Fig1_02062014.eps,width=3.65in}}
\centerline{\includegraphics[width=4.65in]{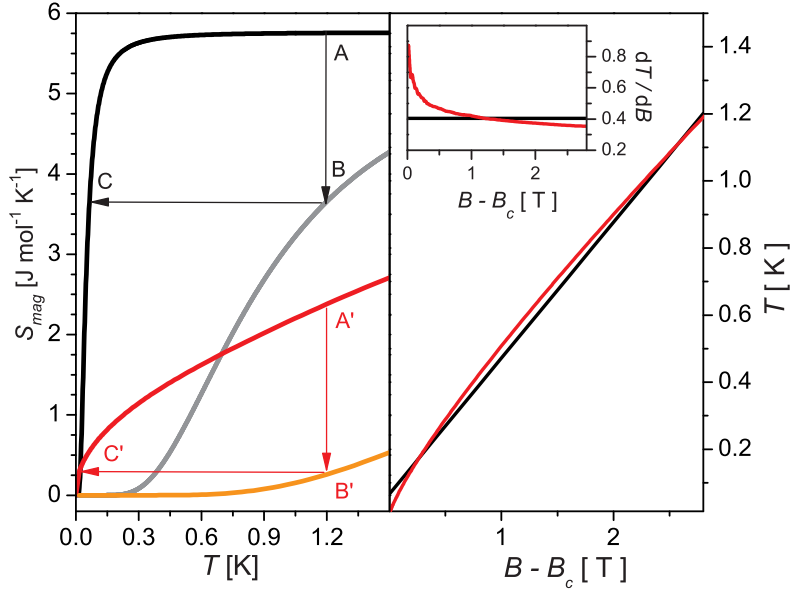}}
%\centerline{\includegraphics[width=3.65in]{test}}
%\vspace*{8pt}
\caption{Left panel: Calculated magnetic entropy $S_{\rm mag}$ of a spin $s$ = 1/2 model paramagnet (PM) with a
zero-field gap of 100\,mK at $B$ = 0\,T (black solid line) and at 2.8\,T (grey line). Also shown is the magnetic
entropy as a function of temperature for an ideal $s$ = 1/2 antiferromagnetic Heisenberg chain (AFHC),
which is inherently quantum critical, with an
intra-chain coupling constant of $J/k_B$ = 3.2~K at two different field values. The red solid line corresponds
to the saturation field of $B_s = 4.1$~T, which marks the endpoint of a quantum-critical line in the
material's $B-T$ phase diagram. The orange line shows the entropy at $B$ = 7\,T where the system is in the field-induced
fully polarized ferromagnetic state. Right panel: cooling curves of the PM (black solid line) and the AFHC (red solid line)
for a field reduction of $\Delta B = -2.8$~T and a starting temperature of 1.2\,K. Note that for the PM the sweep starts
at $B_i$ = 2.8\,T and terminates at $B_f$ = 0 whereas for the AFHC it starts at $B_i$ = 6.9\,T and terminates at
$B_f$ = $B_c$ = 4.09\,T. Inset: Cooling rate d$T$/d$B$ of the PM (black solid line) and the AFHC (red solid line).}
\label{paramagnets}
\end{figure}

Figure \ref{paramagnets} (left panel) shows an adiabatic
demagnetization process for a model paramagnet (black and grey solid
lines) with spin $s = 1/2$ and a zero-field splitting of 0.1\,K. The
process starts with an isothermal magnetization (from A to B) at an
initial temperature $T_i= 1.2$~K, a typical value for a precooling
stage to which the heat of magnetization is released. This process
is followed by an adiabatic demagnetization (from B to C). The black
solid line is the magnetic entropy at zero field $S_{\rm mag}(T, B=
0)$, whereas the grey solid line corresponds to the magnetic entropy
at a finite field, here $B = 2.8$~T. With $T_i = 1.2$~K and a
starting field $B_i = 2.8$~T a final temperature $T_f$ = 0.062\,K is
reached at a final field $B_f$ = 0\,T. The linear variation of
temperature with field, shown by the solid black line in the right
panel of Fig.~\ref{paramagnets}, corresponds to a constant cooling
rate ${\rm d}T/{\rm d}B$ over the whole range of operation (see also
inset of Fig.~\ref{paramagnets}). Whereas the cooling rate d$T$/d$B$
of a paramagnet does not depend on the spin state $s$, the amount of
heat the system is able to absorb during the cooling process does
depend on $s$, see section 4. The typical range of operation for
paramagnets includes magnetic fields varied between a few Tesla and
zero. In their pioneering work Giauque and MacDougall studied the
low-temperature magnetocaloric properties of paramagnetic
Gd$_2$(SO$_4$)$_3$$\cdot$8H$_2$O and demonstrated that a
temperature of 0.53~K could be reached by adiabatic demagnetization.
\cite{Giauque1927} Nowadays paramagnets are mainly used for cooling
to sub-Kelvin temperatures (0.02\,K $\leq T \leq$ 1.2~K), with a
range of operation for a given salt typically stretching over about
one and a half decade in temperature. These systems represent a
suitable alternative to $^{3}$He-$^{4}$He dilution refrigerators --
the present standard technology for continuous operation at $T <1$~K
-- when low costs or ease of operation is in demand. However, they
are of great importance for application at microgravity where standard
dilution refrigerators are not operational. For the status of
gravity-insensitive dilution refrigerators, see
Ref.\,\cite{Camus2014} and references cited therein. Since for space
applications, cooling efficiency is of crucial importance, those
state-of-the-art paramagnetic coolants used there may serve as good
reference materials when discussing performance characteristics of
any new material, see section 4.

\subsection{Quantum-critical systems}

The research on materials in the vicinity of a QCP has been of high
current interest, as quantum criticality, reflecting critical
behaviour arising from quantum instead of thermal fluctuations, may
hold the clue for understanding various intriguing material
properties. \cite{Sachdev2011,Vojta2003} Generally, a QCP can be
reached upon tuning an external parameter $r$ such as pressure,
chemical composition or magnetic field to a critical value. Although
the critical point at $T$ = 0 is inaccessible by experiment, its
presence can affect the material's properties significantly over a
wide range of temperatures. \cite{Sachdev2011,Vojta2003} The
thermodynamic properties of a material in the vicinity of a QCP are
expected to show anomalous power laws as a function of temperature
and, even more spectacular, to exhibit an extraordinarily high
sensitiveness on these tuning parameters. \cite{Zhu2003,Garst2005}
As we have reviewed in section \ref{sec:2}, a diverging magnetic
Gr\"uneisen parameter
$\Gamma_B$ %= -1/$C \cdot$ ($\partial S/\partial B$)$_{T}$
is expected for a $B$-induced QCP,
\cite{Zhu2003,Garst2005,Zhitomirsky2004} and has indeed been
recently observed. \cite{Tokiwa2009,Gegenwart2010,Wolf2011} As has
been first suggested by J. C. Bonner \textit{et al.}
\cite{Bonner1972} and demonstrated by B. Wolf \textit{et al.},
\cite{Wolf2011} this critically enhanced Gr\"{u}neisen parameter
around the QCP can be used in principle for an efficient magnetic cooling.\\

%as shown exemplarily for the spin $s$ = ½ antiferromagnetic Heisenberg chain (AFHC) in the right panel of figure 1?.\\

\subsubsection{Antiferromagnetic spin-1/2 Heisenberg chain}

\begin{figure}[bt]
\centerline{\includegraphics[width=0.3\columnwidth,angle=270]{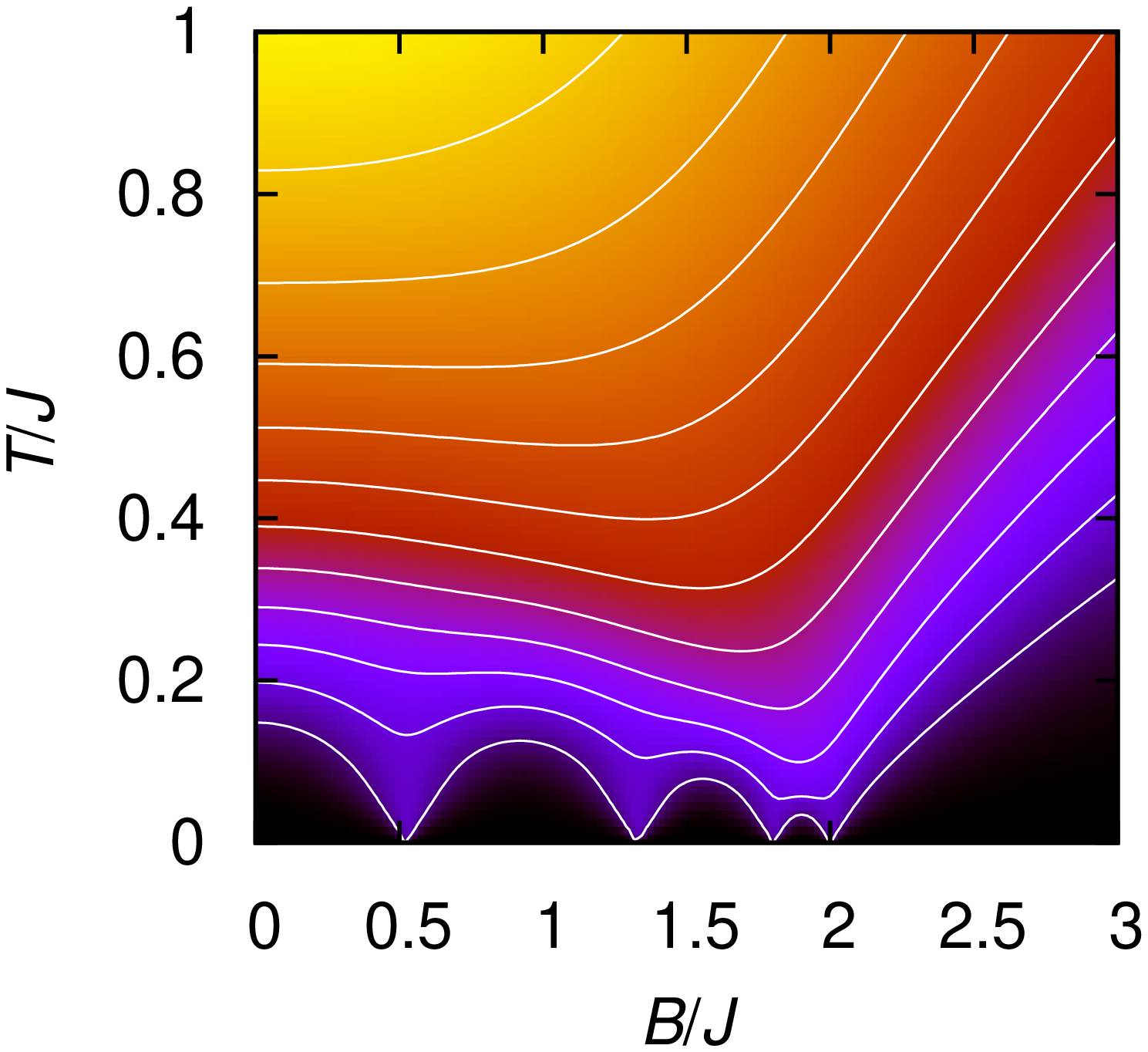}%
\includegraphics[width=0.3\columnwidth,angle=270]{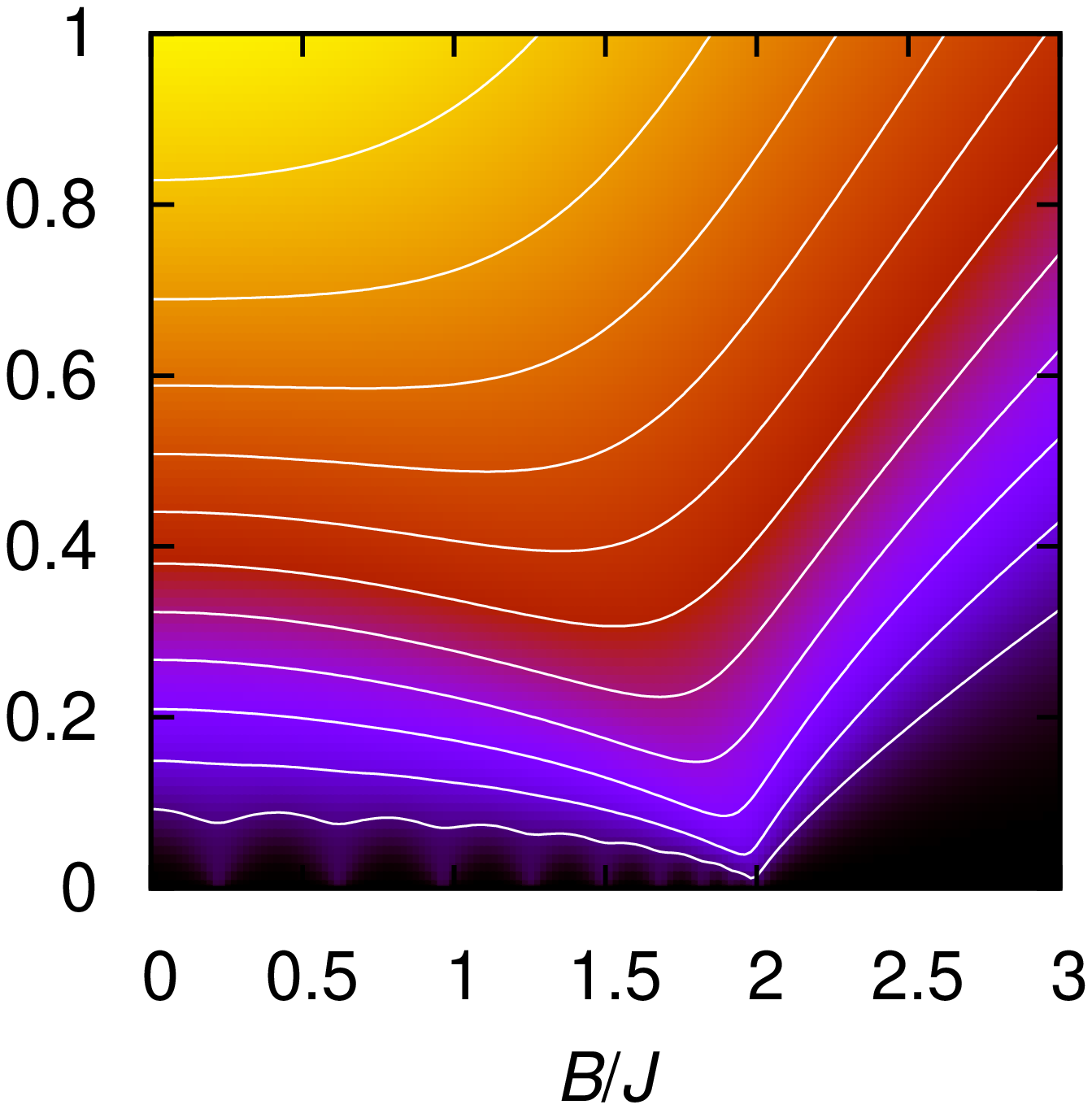}%
\includegraphics[width=0.3\columnwidth,angle=270]{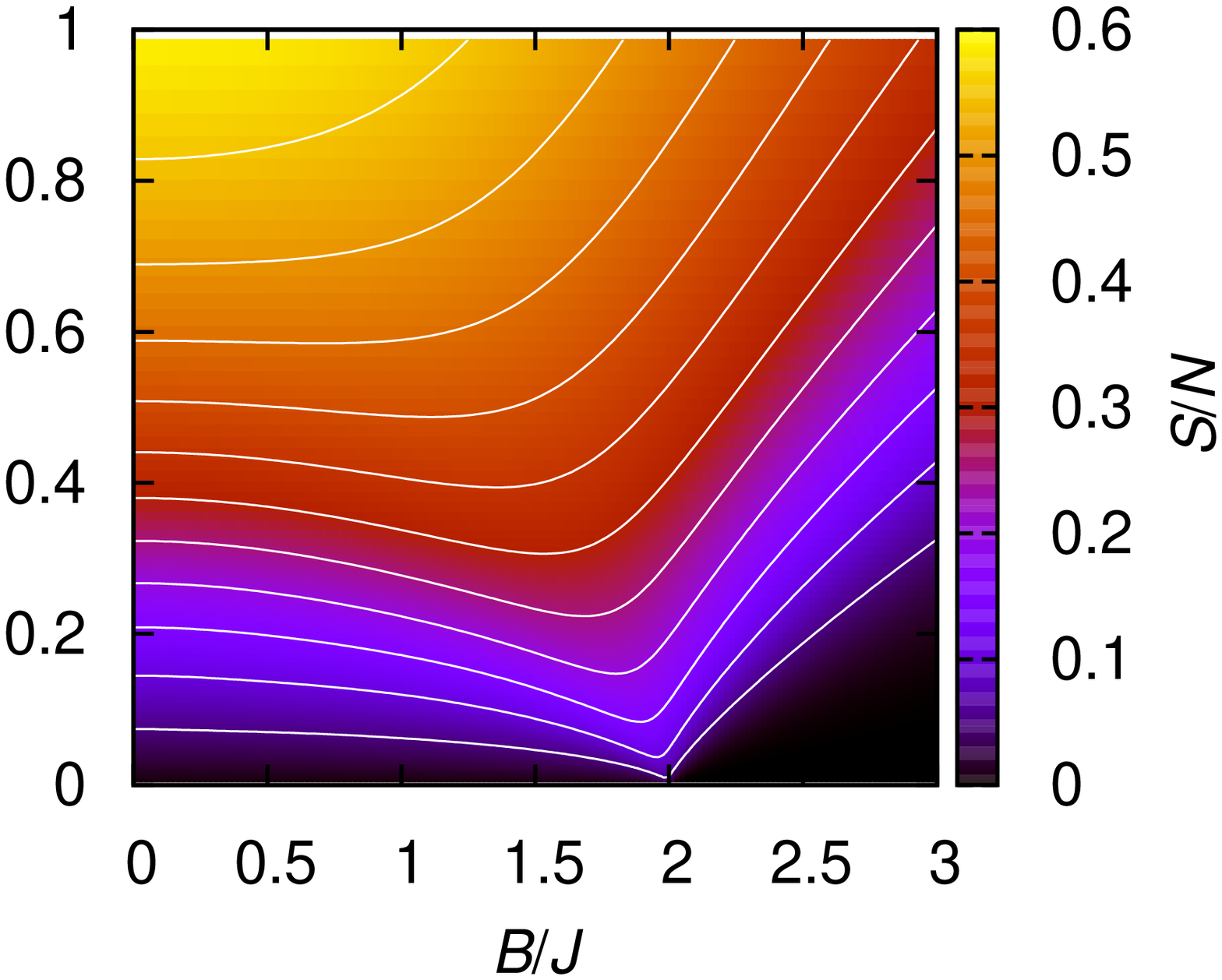}%
}
%\vspace*{8pt}
\caption{Entropy per spin $S/N$ of the uniform spin-1/2 antiferromagnetic Heisenberg chain
as a function of magnetic field $B$ and temperature $T$. The left panel
shows the result for a ring with $N=8$ spins, the middle panel for $N=20$,
and the right panel for the thermodynamic limit $N=\infty$. White contour
lines start at the bottom with an entropy $S/N=0.05$ per spin and increase
in steps of $0.05$. In this figure we use units such that $g\,\mu_B = 1$.
The saturation field $B_s=2\,J$ marks a quantum critical point; for $B<B_s$ the
system is in a gapless Luttinger-Liquid phase whereas for $B > B_s$ it is gapped.
}
\label{fig:SheisChain}
\end{figure}

The uniform spin $s =\frac{1}{2}$ antiferromagnetic Heisenberg chain
(AFHC), where only nearest-neighbour spins at site $i$ and $i+1$
interact via the Heisenberg exchange interaction $J$ along one
crystallographic direction, represents one of the simplest
quantum-critical system. It is described by the Hamiltonian
\begin{equation}
H = J \sum_{i=1}^N \vec{S}_i \cdot \vec{S}_{i+1}
 - g \,\mu_B \,B\, \sum_{i=1}^N S_i^z
\,. \label{Heisenberg}
\end{equation}
Here $J > 0$ is the exchange constant, $S_i$ are spin-1/2 operators at site $i$,
$g$ is spectroscopic splitting factor and $\mu_B$ the Bohr magneton.
This model, despite its simplicity, contains a wealth of non-trivial physics.
The peculiar feature of the AFHC is that its quantum-critical behaviour,
which is that of a Luttinger liquid (LL), governs the material's properties
over wide ranges in temperature and energy (see, e.g., Ref.~\cite{Trippe2010}
for a summary).
The AFHC remains in the LL quantum-critical state for fields up to the
saturation field $g \,\mu_B\,B_s =
%$ \cite{Stone2003}. This field is given by $
2\,J$. \cite{Griffiths1964}
%, with $g$ the spectroscopic $g$-factor and $\mu_B$ the Bohr magneton.
Since the fully polarized state above $B_s$ is a new eigenstate of
the system, different from that of the LL quantum-critical state,
$B_s$ marks the endpoint of a quantum-critical line in the $B-T$
plane. This is illustrated in Fig.~\ref{fig:SheisChain} whose left
panel shows ``historic'' \cite{Bonner1962,Bonner1969} results with
$N=8$, the middle panel $N=20$, \cite{Zhitomirsky2004} and the right
panel the exact solution for the thermodynamic limit $N=\infty$.
\cite{Trippe2010} First, we observe that finite-size effects are
only relevant in the LL regime $B < B_s$ and at low temperatures.
Historically, an important question was if the entropy goes to zero
in the limit $T \to 0$, and it was concluded already on the basis of
the results for $N=8$ that this is indeed the case.
\cite{Bonner1962}

The white lines in Fig.~\ref{fig:SheisChain} show constant entropy curves
(isentropes)
and thus temperature as a function of field during an adiabatic (de)magnetization
process. Indeed, it was recognized early on
by Bonner {\it et al.}\ \cite{Bonner1972,Bonner1977} that
the AFHC exhibits an enhanced MCE and is therefore of interest for being
used for magnetic cooling.
In the vicinity of the QCP at $B=B_s$ magnetic entropy is accumulated
as a consequence of the competition between the different ground states on both sides.
This leads to a minimum of the isentropes above this QCP, i.e., a cooling
effect upon approaching this QCP.
In order to give a first impression on some of the cooling characteristics
of such a quantum-critical system, we show in Fig.~\ref{paramagnets} the
relevant quantities of an AFHC in comparison with those of a paramagnet.
The cooling performance of a good realization of an AFHC will be discussed
in detail in subsection \ref{sec:Qcrit}. As Fig.~\ref{paramagnets} illustrates,
quantum-critical systems are of particular interest when an efficient
cooling to (in principle) arbitrarily low temperatures is in demand.

\subsubsection{Dimerized spin-1/2 Heisenberg chain}

\begin{figure}[bt]
\centerline{\includegraphics[width=0.4\columnwidth,angle=270]{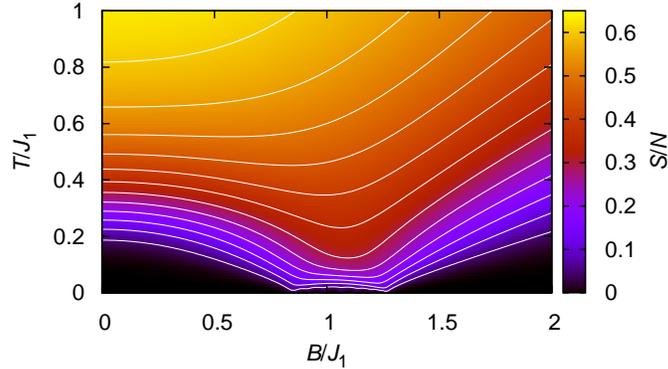}}
%\vspace*{8pt}
\caption{Entropy per spin $S/N$ of a dimerized spin-1/2 Heisenberg chain
with $J_2 = 0.27\,J_1$ and $N=20$ spins. White contour
lines start at the bottom with an entropy $S/N=0.05$ per spin and increase
in steps of $0.05$. In this figure we use units such that $g\,\mu_B = 1$.
In this case, there are two quantum critical points: one at
the saturation field $B_s=1.27\,J_1$ and one at
$B_c \approx 0.838\,J_1$. For $B_c<B<B_s$ the
system is in a gapless Luttinger-Liquid phase whereas for $B<B_c$ or
$B > B_s$ it is gapped.
}
\label{fig:SdimChain}
\end{figure}

Let us briefly review a second prototypical system, namely a
``dimerized'' $s=1/2$ Heisenberg chain where two different
exchange constants $J_1$ and $J_2$ alternate. Historically, MCE
measurements on copper nitrate (Cu(NO$_3$)$_2\cdot$2.5H$_2$O)
\cite{Amaya1969,vanTol1973} prompted theoretical investigations of
the MCE in the corresponding spin-1/2 dimer chain model
\cite{Tachiki1970a,Tachiki1970b,Diederix1979}. Figure
\ref{fig:SdimChain} shows the entropy as a function of field and
temperature for the historical parameters $J_2 = 0.27\,J_1$.
\cite{Diederix1979} Although this is a chain with 20 spins (10
dimers), the strong dimerization $J_2/J_1 = 0.27$ ensures that
finite-size effects are neglibile except for $g\,\mu_B\,B$
comparable to $J$ and very low temperatures. In this case, there are
two QCPs: one at the saturation field $g\,\mu_B\,B_s = J_1 + J_2 =
1.27\,J_1$ and another one at $g\,\mu_B\,B_c \approx 0.838\,J_1$.
The system is gapped for $B<B_c$ or for $B>B_s$ and in a gapless LL
state for $B_c < B < B_s$. Accordingly, one finds a strong cooling
effect for adiabatic field variations $B \nearrow B_c$ or $B
\searrow B_s$ and comparably small temperature variations in the LL
state.

Recently, the exchange couplings of copper nitrate have been refined
using inelastic neutron scattering \cite{Tennant2012} and there are
new experimental results for the MCE available (see, e.g.,
Ref.~\cite{Willenberg2014}). One challenge that remains is to
provide a quantitative description close to the ordering transition
of copper nitrate at $T \lesssim 150$\,mK \cite{Willenberg2014}
which is beyond the simple dimerized chain model.
\cite{Diederix1979} The ordering transition is captured
qualitatively by a dimer mean-field theory,
\cite{Tachiki1970a,Tachiki1970b} but this theory is not appropriate
for describing the LL state. A chain mean-field theory might provide
a reasonable alternative for describing the MCE of the coupled-chain
system, as has been demonstrated for similar situations.
\cite{Lang2013}

\subsubsection{A highly frustrated magnet: spin-1/2 Heisenberg model on the kagome lattice}

\label{sec:kag}

\begin{figure}[bt]
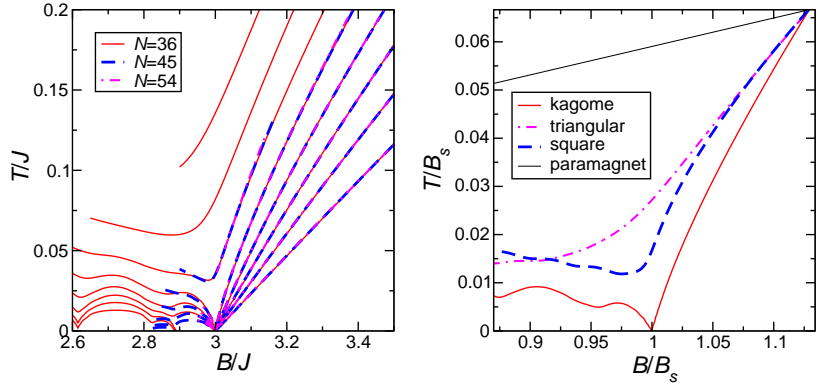

\centerline{\includegraphics[height=0.4\columnwidth]{wolf_Fig5a.eps}\quad\includegraphics[height=0.4\columnwidth]{wolf_Fig5b.eps}}
%\vspace*{8pt}
\caption{Left panel: Isentropes for the spin-1/2 Heisenberg model on the kagome lattice
close to the saturation field $B_s = 3\,J$.
Lines start at the bottom with an entropy $S/N=0.025$ per spin and increase
in steps of $0.025$.
Right panel: Comparison of an adiabatic demagnetization process of
a spin-1/2 Heisenberg model on the kagome, triangular, and square lattice.
The system size is $N=36$ in all three cases.
Also included is the straight line characteristic of an ideal paramagnet.
In both panels we use units such that $g\,\mu_B = 1$.
}
\label{fig:kagent}
\end{figure}

Lastly we illustrate the effect of geometric frustration using the spin-1/2 Heisenberg model on
a prototypical frustrated lattice: the kagome lattice. The left panel of Fig.~\ref{fig:kagent} shows exact diagonalization
results for different system sizes close to the saturation field $g \,\mu_B\,B_s = 3\,J$.
Although some extremely big matrices haven been diagonalized in this context,  systems with $N \ge 36$ spins
are too big to allow for a full diagonalization. Therefore, we have performed a low-energy approximation
using low-energy states. This in turn implies that one has to restrict the field range of some isentropes in order to render
truncation effects small, compare also Refs.~\cite{Honecker2009,Honecker2006}. Comparison of the different system
sizes shown in the left panel of Fig.~\ref{fig:kagent} shows that finite-size effects are small, at least for
$B \gtrsim B_s$. This reveals a remarkable property: the third law of thermodynamics is violated in this model --
there is a finite zero-temperature entropy at the saturation field $B=B_s$. Accordingly, any adiabatic process starting
with an entropy less than this zero-temperature entropy necessarily goes to $T=0$ as $B \to B_s$.
With the help of the localized magnon ground-states one can give lower bounds for the zero-temperature
entropy \cite{Schulenburg2002,Zhitomirsky2004,Derzhko2007} and show rigorously that it is indeed finite.
Inspection of the ground-state degeneracies \cite{Derzhko2007} corresponding to the data in the left panel of Fig.~\ref{fig:kagent}
yields the estimate $S/N \approx 0.13$ for the zero-temperature entropy, amounting to about $18\%$ of the total
entropy of the system. However, for $B < B_s$, the ground-state degeneracy of the classical system is lifted
by quantum fluctuations and one obtains a certain amount of heating if one decreases the magnetic field further below
$B_s$.

In order to illustrate the importance of geometric frustration, we imagine the following thought
experiment: we take the non-frustrated square lattice as well as the frustrated triangular lattice
and normalize the energy scale such that the quantum critical points at $B = B_s$ coincide.
Then we start at one point in the $(B,T)$-plane and compare the adiabatic demagnetization processes
of the different systems. This is implemented in the right panel of Fig.~\ref{fig:kagent} which
shows one isentrope each for the spin-1/2 Heisenberg model on the square, triangular, \cite{Honecker2006}
and kagome lattice, using systems with $N=36$ spins in all three cases. We also include the straight line characteristic for an ideal paramagnet. Evidently,
any quantum critical system outperforms the ideal paramagnet in this field range and the highly frustrated
kagome lattice outperforms the other systems in the sense that one can theoretically reach arbitrarily low temperatures.
The right panel of Fig.~\ref{fig:kagent} suggests that the non-frustrated square lattice outperforms the frustrated triangular
lattice in the sense that a lower temperature is reached close to the saturation field for these particular initial conditions.
Nevertheless, other aspects also need to be kept in mind. In particular, the entropy per spin along these isentropes
is $S/N = 0.0258$, $0.067$, and $0.1$ for the square, triangular, and kagome lattice, respectively. Thus, with increasing
frustration, the amount of entropy increases that is available for the exchange of heat with  a cooling load.

In any real system, the third law of thermodynamics will of course be restored, thus limiting the smallest
temperature that one can reach. Still, one may hope some of the good properties of a highly frustrated magnet
to be preserved in a relevant temperature range.

\subsection{Interacting ultracold gases}
\label{sec:cold_gases}

Large theoretical and experimental efforts have been devoted to
revealing the mechanisms behind magnetic ordering and
superconductivity of quantum many-body systems. Due to the high
level of complexity in solid state systems, a quantitative
comparison between microscopic theory and experiment can be a
challenging task. It is therefore highly desirable to work with
systems that are able to simulate the original solid state many-body
systems, but with a higher degree of tunability and control.

Over the past decade, ultracold quantum gases in optical lattices
have provided an excellent laboratory for investigating quantum
many-body physics. \cite{Jaksch1998,Bloch2008} Optical lattices of
different geometry (triangular, hexagonal, kagome) have been
created, \cite{Soltan-Panahi2011,Jo2012} and Mott insulating phases
(bosonic, fermionic) realized, \cite{Joerdens2008,Schneider2008} to
mention just a few of the relevant developments. Correlated atom
tunneling and superexchange couplings have been observed,
\cite{Trotzky2008} as well as short-range magnetic correlations.
\cite{Greif2013} New single-atom in-situ detection techniques with
single-site resolution, based on the quantum-gas microscope,
\cite{QuantumGasMicroscope} allow the detection of magnetic
excitations and their dynamics. \cite{Fukuhara2013} A quantum
simulation of the Ising model in a transverse magnetic field and its
ordering transition have been achieved via mapping to a tilted Mott
insulator and single-site resolved detection. \cite{Simon2011} One
major current goal of the field is to simulate the phase diagram of the
fermionic Hubbard model, including antiferromagnetic long-range
magnetic order. \cite{Hofstetter2002} More generally, it has been
shown that also bosonic two-component gases allow for tunable
(ferro- and antiferromagnetic) spin order.
\cite{Duan2003,Altman2003,Hubener2009} The corresponding critical
temperatures are defined by the superexchange scale and have not
been reached yet in experiments. \cite{Joerdens2010} While \emph{average} entropies
per particle down to $S / k_B N = 0.5 $ have been demonstrated for
fermions in pure dipole traps, in optical lattices only values down
to $S / k_B N = O(1) $ could be realized so far. \cite{McKay2011}
Note that in most cases the entropy density in the center of the
optical trap is significantly lower. Nevertheless, it is so far
above the critical entropy per particle for long-range magnetic order in a
lattice, which for the fermionic case (N\'{e}el order) has been
calculated as $S_c/N \approx 0.5 k_B \ln 2 = 0.35 k_B $,
\cite{Werner2005} while for two--component bosons in $d=3$ it has been
shown that the critical entropy depends on the type of long-range
order: it is given by the value $S_c/N \approx 0.5 k_B$ for the z-N\'{e}el
state (where only a discrete Ising-symmetry is broken) and the lower
value $S_c/N \approx 0.35 k_B $ for the xy-ferromagnet (where,
analogously to spin-$1/2$ fermions, a continuous symmetry is broken,
in this case $U(1)$). \cite{Capogrosso-Sansone2010}

In order to reach these low entropies, which are characteristic for
strongly correlated states with magnetic order, new cooling
techniques in the presence of the optical lattice need to be
implemented. Note that direct evaporative cooling -- the standard
approach to degenerate quantum gases -- is in general not efficient
in the lattice due to low scattering rates. \cite{McKay2011} A
further challenge in this context is the \emph{dynamical arrest} of
many-body dynamics, which was recently discovered for a strongly
interacting fermionic mixture. It leads to departure from adiabatic
evolution of the atomic cloud when the lattice is ramped up,
\cite{Schmidt2013} and therefore implies a higher entropy in the
final state. To avoid this effect, cooling schemes which avoid
global mass- or entropy transport could be preferable.
\cite{Lubasch2011}

Inspired by solid state systems, adiabatic demagnetization cooling
of an ultracold gas has been performed, \cite{Fattori06} with final
temperatures in the $\mu$K range. However, so far this technique has
only been applied for $^{52}$Cr, which has a high magnetic moment
and therefore a high dipolar relaxation rate. \cite{Fattori06} This
approach has not yet been implemented in combination with an optical
lattice.

\begin{figure}[bt]
%\centerline{\psfig{file=Fig2_02062014.eps,width=3.65in}}
\centerline{\includegraphics[width=2.4in]{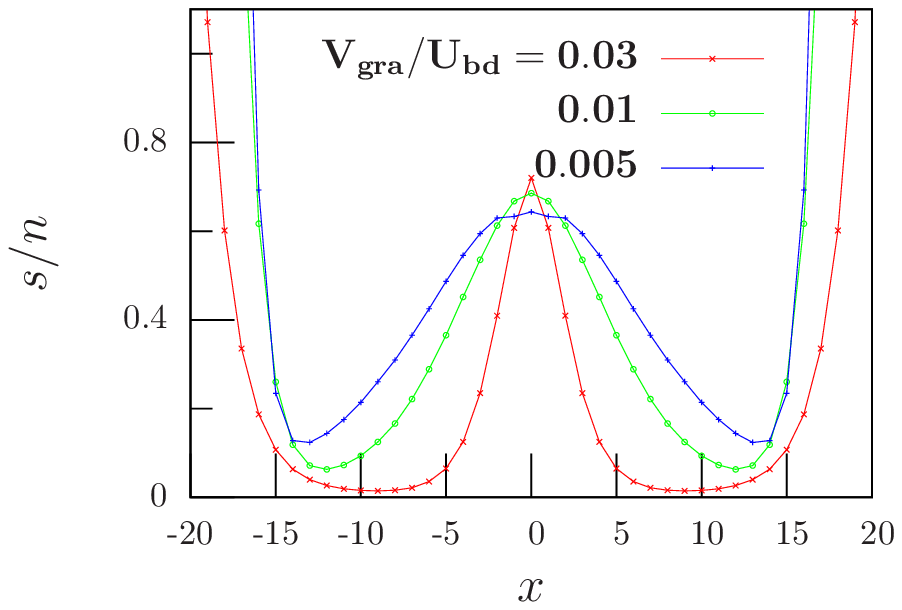} \
\includegraphics[width=2.4in]{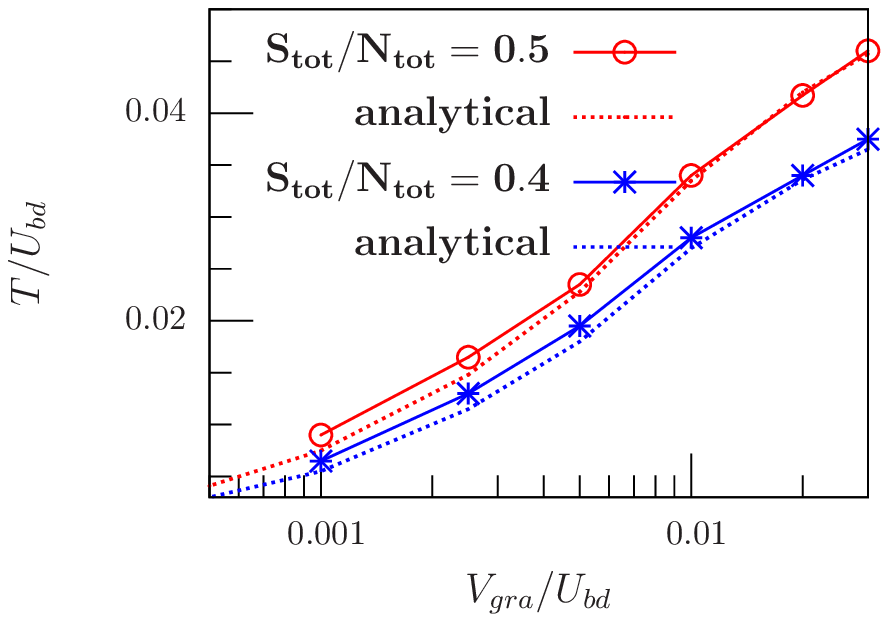}  }
%\vspace*{8pt}
\caption{Spin-gradient demagnetization cooling.
Left: Field-gradient dependence of the distribution of the local entropy per particle
along the $x$ direction on the $y,z=0$ axis, for an
average entropy per particle $S_{tot}/N_{tot}=0.7$. The red, green and
blue lines correspond to field gradient values of $V_{gra}/U_{bd}=0.03$,
$0.01$ and $0.005$.
Right: Adiabatic spin-gradient demagnetization cooling
in a cubic lattice. Results are obtained by BDMFT+LDA and compared to an analytical
zero-tunneling approximation in the atomic limit.
\protect\cite{Weld2010}
Interaction parameters
are set to $U_b=U_d=1.01U_{bd}$, and the hopping amplitudes are given by
$2zt_b=2zt_d=0.12U_{bd}$, with a total particle number $N_{tot}
\approx 17000$ in a harmonic trap
$V_0=0.0025U_{bd}$.
Figures taken from \protect\cite{Li2012}.
\label{spin_gradient_cooling}
}
\end{figure}

On the other hand, a related technique -- spin-gradient thermometry
and cooling -- has been realized for a mixture of two hyperfine
states of $^{87}$Rb in a 3D optical lattice,
\cite{Weld2009,Medley2011} and theoretically simulated.
\cite{Li2012} Here, the ultracold Bose gas is prepared in a magnetic
field gradient, which separates the two spin components to opposite
sides of the trap. The width of the transition layer between two
hyperfine spin domains serves as an accurate thermometer for
temperatures in the nK regime, and entropies in a broad range $0.1 <
S/N k_B < \ln 2 $. As the gradient is lowered, the two spin
components mix, and entropy from particle-hole excitations is
transferred into the spin mixing entropy. Using bosonic dynamical
mean-field theory (DMFT), we have quantitatively verified that this
leads to a cooling effect for the interacting many-body system, and
shown that the magnetically ordered regime (XY ferromagnet) can be
reached in this way, \cite{Li2012,Li2011} see
Fig.~\ref{spin_gradient_cooling}.
%To be specific, a Maxwell relation $\partial s / \partial n = \partial n / \partial T$ allows us to determine the entropy distribution in the trap
%by DMFT in combination with a local-density approximation.
In the following we give more details.
The Hamiltonian of the two--component bosonic $^{87}$Rb system is given by
\begin{eqnarray} \label{Hamil}
\mathcal{H}=&-& \sum_{\stackrel{<i,j>}{\nu=b,d}} t_\nu (b^\dagger_{i\nu}b_{j\nu}+h.c.)+\frac{1}{2}\sum_{i,\lambda\nu} U_{\lambda\nu} \hat{n}_{i\lambda}
(\hat{n}_{i\nu}-\delta_{\lambda\nu}\nonumber) \\
&+&\sum_{i,\nu=b,d} (V_i-\mu_\nu)\hat{n}_{i\nu} - \sum_{i,\nu}
\mu^\nu_{mag}B(x_i)\hat{n}_{i\nu}
\end{eqnarray}
where we include a linear position-dependence of the magnetic field
in $x$ direction, $B(x_i)=c\, x_i$ where $c$ is the magnetic field
gradient and $x_i$ the distance from the harmonic trap center, as in
the experiment. \cite{Medley2011}

Here $\langle i,j\rangle$ denotes the summation over nearest
neighbours sites and the two bosonic species (hyperfine states) are
labeled by the index $\lambda (\nu)=b,d$. These two species can in
general have different hopping amplitudes $t_b$ and $t_d$, due to a
spin-dependent optical lattice. We denote the bosonic creation
(annihilation) operator for species $\nu$ at site $i$ by
$b^{\dagger}_{i\nu}$ ($b_{i\nu}$) and the local density by
$\hat{n}_{i,\nu}=b^\dagger_{i\nu}b_{i\nu}$. $U_{\lambda\nu}$ are the
inter- and intra-species interactions, which are tunable via
Feshbach resonances or by a spin-dependent lattice. For $^{87}$Rb we
assume the couplings $U_{bb} \approx U_{dd} \approx U_{bd}$ to be approximately
equal, which was the case in the experiment. \cite{Medley2011}
$\mu_\nu$ denotes the global chemical potential for the two bosonic
species, while $V_i$ is the harmonic potential due to the optical
trap, and $\mu^\nu_{mag}$ is the magnetic moment of component $\nu$.
\begin{figure}[bt]
%\centerline{\psfig{file=Fig2_02062014.eps,width=3.65in}}
\centerline{\includegraphics[width=3.0in]{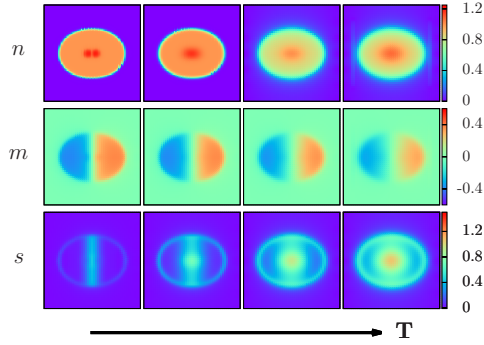}  }
%\vspace*{8pt}
\caption{
Real-space distribution (in the $z=0$ plane of a 3D cubic optical lattice) of the local particle density $n$, magnetization $m$ and entropy $s$,
calculated by BDMFT+LDA. From left to right, the temperature is increasing as $T/U_{bd}=0.020$, $0.040$, $0.070$ and $0.095$, respectively. The interactions are $U_b=U_d=1.01U_{bd}$ and the hopping amplitudes are set to $2zt_b=2zt_d=0.12U_{bd}$, with a total particle number $N_{tot}\approx 17000$ in a harmonic trap of strength $V_0=0.004U_{bd}$ and a magnetic field gradient $V_{gra}=0.01U_{bd}$.
Figure taken from \protect\cite{Li2012}.
\label{entropy_distribution}
}
\end{figure}
Bosonic dynamical mean-field theory (BDMFT)
\cite{Byczuk2008,Hubener2009,Hu2009,Anders2010,Snoek2013} provides a
non-perturbative description of strongly correlated bosons on a
lattice at zero- and finite-temperature. We have recently developed
a \emph{real-space} version of BDMFT (RBDMFT), \cite{Li2011} in
order to take into account inhomogeneities such as an external
trapping potential or disorder. The results presented here have,
however, been obtained by a local density approximation (LDA) scheme
in combination with single-site BDMFT. The advantage of this
approach lies in the larger system size accessible. In these
calculations, the local chemical potential is adjusted according to
the trapping potential, i.e., $\mu_\nu(r)=\mu_\nu-V_0r^2$, where
$V_0$ characterizes the strength of the harmonic confinement and $r$
is the distance from the center of the trap.We have benchmarked the
validity of the LDA+BDMFT approach by a quantitative comparison with
the more rigorous RBDMFT method.

A direct calculation of the entropy within BDMFT is problematic. But, assuming that the system is in thermodynamic equilibrium,
we can apply a Maxwell relation to obtain the local
entropy per site at temperature $T$ and chemical
potential $\mu_s(r)=(\mu_b(r)+\mu_d(r))/2$:
\begin{equation}\label{entropy}
s(\mu_s(r_0),T)=\int_{-\infty}^{\mu_s(r_0)} \frac {\partial n
(r)}{\partial T}d\mu_s (r)
\end{equation}
where $n (r)=n_b+n_d$ denotes the local density ({\it i.e.}, number of
particles per lattice site) at a distance $r$ from the trap center. Note that the thermodynamic relation
(\ref{entropy}) is only valid at fixed difference of the chemical potentials $\Delta \mu(r)=
\mu_b(r)-\mu_d(r)$ for the two-component mixture.
(R-)BDMFT yields density distributions which are accurate enough to allow for a precise evaluation
of the derivative $\frac {\partial n}{\partial T}$.
The relation (\ref{entropy}) has been used to obtain the entropy distributions which will be discussed in the following.

The spin-gradient cooling scheme is based on the inhomogeneous
entropy distribution of the trapped system, as shown in
Fig.~\ref{entropy_distribution}. During the demagnetization process,
the local entropy per particle in the spin-mixed regions decreases.
For the fillings considered here, three different spatial regions
correspond to different phases of the system: the superfluid, the
spin-mixed and the single-component Mott-insulating regions. For
large initial values of the field gradient, the superfluid and the
spin-mixed region carry most of the entropy of the system, while the
entropy in the single-component Mott insulator regions is
negligible. With decreasing magnetic field gradient, the spin-mixed
region expands, while the single-component Mott-insulating region
shrinks. At the same time, the temperature drops, although entropy
carried by hot mobile particles is drained into the expanding mixed
region, since the local entropy \emph{per particle} in the central
spin-mixed region decreases. In the left panel of
Fig.~\ref{spin_gradient_cooling}, we show the local entropy per
particle $s/n$ at different field gradient strengths for fixed total
particle number and entropy. It is clearly visible that $s/n$
decreases in the central region as the field gradient is
adiabatically reduced. As a result, the temperature decreases from
$T/U_{bd}=0.065$ to $0.035$ when the field gradient is reduced from
$V_{gra}/U_{bd}=0.03$ to $0.005$. The resulting spin gradient
demagnetization cooling efficiency is displayed in the right panel
of Fig. \ref{spin_gradient_cooling}. Our result from BDMFT
simulations is in good agreement with analytic results from the
zero-tunneling (atomic) limit, \cite{Weld2010} since in our
description of the experiment we choose a deep optical lattice which
leads to a small hopping $t_\nu/U_{bd}$. The adiabatic
demagnetization cooling appears to be less efficient at larger
values of the field gradient, since in this case the trap center
becomes superfluid, with enhanced entropy capacity.

Spin-gradient cooling efficiency may be additionally limited by
magnetic correlations when the temperature is of the order of the
super-exchange energy, \cite{Medley2011} and (dynamically) by the
fact that it requires global entropy transport. The latter problem
is avoided by proposals based, e.g., on superlattices,
\cite{Lubasch2011} which require only local mass and entropy
transport on length scales of the order of the lattice constant, and
may thus allow creating N\'eel-ordered states, starting from a
low-entropy band insulator.

\begin{figure}[bt]
%\centerline{\psfig{file=Fig2_02062014.eps,width=3.65in}}
\centerline{\includegraphics[width=3in]{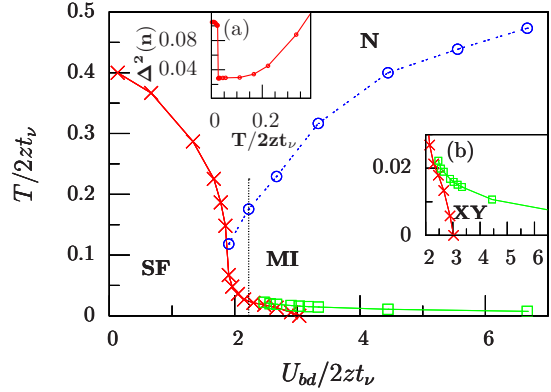}}
%\vspace*{8pt}
\caption{Pomeranchuk effect in spinful ultracold bosons.
Upon heating of the strongly correlated, non-ordered superfluid (SF) phase
along the dotted line shown
the system enters into the paramagnetic Mott insulator (MI), which has the higher spin entropy $S/N \approx k_B \ln 2$.
This results in a drastic reduction and non-monotonic temperature-dependence of local number fluctuations $\Delta(n_i)$, which could be measured
in-situ with the optical quantum-gase microscope,
\protect\cite{QuantumGasMicroscope}
and occurs at higher temperatures/entropies than the
transition into the long-range ordered phase with xy-ferromagnetism (below the green line).
Figure taken from \protect\cite{Li2012}.
%** \emph{hier noch Bild einfuegen}** To the right: the corresponding effect in the double occupancy for the spin-1/2 fermions, as predicted by \protect\cite{Werner2005}
\label{Pomeranchuk}}
\end{figure}

\begin{figure}[bt]
%\centerline{\psfig{file=Fig2_02062014.eps,width=3.65in}}
\centerline{\includegraphics[width=3.65in]{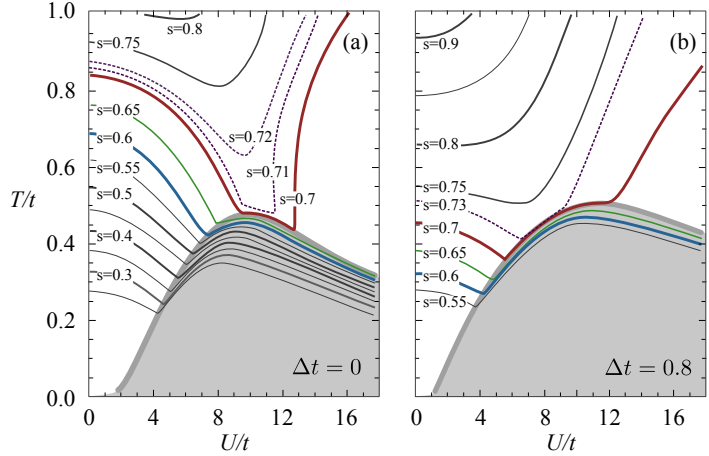}}
%\vspace*{8pt}
\caption{Pomeranchuk cooling of ultracold fermions with spin-$1/2$. Preparing a low-entropy degenerate mixture,
adiabatic loading into a shallow lattice and ramp-up of the lattice creates a N\'eel-ordered state (grey region).
Note that the critical entropy is overestimated by DMFT as $S/N \approx k_B \ln2$. This DMFT estimate becomes more accurate
for the hopping-imbalanced system b).
Figure taken from \protect\cite{Sotnikov2012}.}
%\cite{Sotnikov}.
\label{Pomeranchuk_cooling}

\end{figure}

A further many-body phenomenon leading to interaction-enhanced
cooling is the \emph{Pomeranchuk effect}, which was first discovered
in $^3$He, where heating leads to increased localization due to the
higher (nuclear) spin entropy of the solid phase. An analogous
phenomenon occurs within the Hubbard model, where at intermediate
coupling strengths the spin entropy of the Mott insulator $S/N
\approx k_B \ln 2$ is higher than that of the Fermi liquid.
\cite{Werner2005} Recently a similar effect was discovered for
spinful bosonic mixtures, e.g., of two hyperfine states of $^{87}$Rb,
where upon heating of the strongly correlated, non-ordered
superfluid phase (Fig.~\ref{Pomeranchuk}, dotted line) the system
enters into the paramagnetic Mott insulator, which again has the
higher spin entropy $S/N \approx k_B \ln 2$. \cite{Li2012} This
results in a drastic reduction and non-monotonic temperature
dependence of the local particle number fluctuations $\Delta^2(n_i)
= \langle \hat{n}^2_i \rangle - \langle \hat{n}_i \rangle^2$ on a
single site, which could be measured in-situ with the optical
quantum-gas microscope. \cite{QuantumGasMicroscope} Since this
effect occurs at temperatures and entropies above the magnetic
ordering scale, it may be more easily accessible experimentally than
long-range magnetic order. The Pomeranchuk effect also leads to
enhaced adiabatic cooling beyond noninteracting band structure
effects. \cite{Hofstetter2002,Werner2005}
In Fig. \ref{Pomeranchuk_cooling}
we show results for the spin-$1/2$ Fermi-Hubbard model, including
mass (hopping) imbalance. \cite{Sotnikov2012,Sotnikov2013}
Preparing a low-entropy degenerate mixture, followed by adiabatic
loading into a shallow lattice and ramp-up of the lattice, leads to a
N\'eel-ordered state (grey region). Note that the critical entropy
is overestimated by DMFT as $S_c/N \approx k_B \ln2$, as is the
magnitude of the Pomeranchuk effect, which for fermions has been
shown to decrease due to short-range magnetic correlations not
captured by DMFT. \cite{Fuchs2011} The DMFT estimate for $S_c$
becomes, however, more accurate for the hopping-imbalanced system,
where the full $SU(2)$ symmetry is broken down to $Z_2 \times U(1)$,
which leads to a gapped excitation spectrum and therefore
constitutes a promising route towards superexchange-mediated quantum
magnetism. The Pomeranchuk effect has been demonstrated
experimentally for ultracold fermionic $^{173}$Yb mixtures with
internal symmetry $SU(N)$ up to $N=6$, where it has been shown that
the corresponding larger value of the (hyperfine) spin enhances the
magnitude of the adiabatic cooling effect and leads to lower
temperatures in the final Mott-insulating state. \cite{Taie2012a} In
a related work on high-spin fermions, it has been pointed out that
inhomogeneous entropy profiles in the presence of an optical trap
and a spatially variable quadratic Zeeman coupling may lead to
highly efficient adiabatic cooling of an effective spin-$1/2$ core,
thus leading to antiferromagnetic order. \cite{Colome-Tatche2011}

%\begin{itemize}
%\item known from $^3$He
%\item increasing localization with heating, due to higher spin entropy $S/N \approx k_B \ln 2$ of the (Mott-)localized state
%\item leads to non-monotonic dependence of $d$ and $\Delta(n_i)$ as a function of temperature,
%directly observable by optical quantum gas microscope
%\item has been predicted for fermionic Hubbard model [Werner05] and more recently for bosons [Li]
%\item weakened by non-local magnetic correlations beyond DMFT
%\item leads to enhanced cooling upon lattice ramp-up [Werner05, Sotnikov]
%\item beyond simple (non-interacting) adiabatic cooling of fermions due to band structure [Hofstetter02]
%\item idea: prepare low-entropy state, then ramp up lattice into long-range ordered state
%\item critical entropy can be increased by mass (hopping) imbalance -> lower symmetry, XXZ type spin model
%or Falicov-Kimball model [Sotnikov]
%\item magnitude of effect increases for higher symmetry $SU(N)$ [Rey, Hazzard, Takahashi, Santos?] -> enhanced cooling
%\end{itemize}

\subsection{Magnetic cooling of solid state systems}

\subsubsection{State-of-the-art paramagnets}

Similar to the classical coolant Gd$_2$(SO$_4$)$_3$$\cdot$8H$_2$O,
dilute systems of magnetic ions are used for technical applications.
The standard materials are the spin-5/2 containing ferric ammonium
alum Fe(NH$_4$)(SO$_4$)$_2 \cdot$12H$_2$O (FAA), \cite{Cooke1949}
the spin-3/2 carrying chromic potassium alum CrK(SO$_4$)$_2
\cdot$12H$_2$O (CPA), \cite{Bleaney1950} and the effective spin-1/2
compound cerium magnesium nitrate (CMN). \cite{Cooke1953} These
materials have in common that they all contain a significant amount
of water to dilute the magnetic system. Despite the large distance
between the magnetic ions, precluding their direct magnetic exchange
interaction, there are still weak residual magnetic interactions
which eventually leads to long-range magnetic order in all of these
compounds at a critical temperature $T_c \leq$ 50\,mK. The magnetic
order leads to a significant reduction of $S_{\rm mag}$, and
therefore $T_c$ marks the lower bound for the accessible temperature
range. In addition, on approaching $T_c$ from above, the cooling
becomes progressively inefficient. Due to the reduced entropy (as a
consequence of incipient short-range ordering), the adiabatic
temperature change is smaller than for an ideal paramagnet. In a
standard  single-stage magnetic cooler the typical field range for
the demagnetization process is $0 \leq B \leq 2$~T with an initial
temperature $T_i \leq 1.5$~K. \cite{Pobell1992} The material CMN,
having a $T_c$ of about 2\,mK, can be used as a magnetic coolant
down to temperatures even below 5\,mK. This has to be compared with
the lowest temperature of around 15\,mK for CPA and 30\,mK for FAA.
Since the amount of heat which can be absorbed by a paramagnetic
coolant depends on the spin $s$, FAA ($s = 5/2$) is often used when
a large amount of heat has to be removed from the sample. This is
the case in some space applications. \cite{Shirron2007,Hagemann1999}
Due to its entropy distribution as a function of temperature and
field, FAA is a practical working substance for a single-stage
magnetic cooler. But unfortunately, FAA has a few properties which
complicate its use. Among them is the fact that it is corrosive to
copper-based alloys. In addition, FAA expands slightly when crystals
are formed and undergoes a chemical transition when heated above
40$^\circ$~C which permanently destroys its paramagnetism. Moreover,
FAA dehydrates when exposed to vacuum and the resulting powdery
substance does no longer act as an effective coolant at low
temperatures. This means that special materials together with
peculiar techniques are necessary to build the housing of a FAA salt
pill. \cite{Wilson1995} Another problem in connection with all
paramagnetic salts is their low thermal conductivity. This problem
is usually circumvented by putting some Au-wires inside the salt
pill which provide a sufficient thermal contact between the
paramagnetic salt and the sample/device to be cooled. An alternative
approach for tackling the problem of the low thermal conductivity of
paramagnetic salts is the use of paramagnetic intermetallic
compounds. They have attracted some attention with respect to their
magnetocaloric properties. One of the most studied materials in this
context is PrNi$_5$, which has also been
successfully used in nuclear adiabatic demagnetization devices. \cite{Mueller1980}\\

\subsubsection{Frustrated magnets: classical systems}

An enhanced cooling power in comparison to paramagnets can be
obtained with magnetic materials exhibiting geometric magnetic
frustration because these materials stay in the so-called disordered
cooperative paramagnetic state well below the Curie-Weiss
temperature. \cite{Villain1971} Due to their large spin values these
materials are labeled as classical systems. A prominent example is
Gadolinium gallium garnet Gd$_3$Ga$_5$O$_{12}$ which has long been
known as a suitable refrigerant material below 5\,K. Effective
cooling has been demonstrated down to temperatures of about 0.5\,K.
\cite{Barclay1982} Another group of geometrically frustrated magnets
beside the garnets are Heisenberg antiferromagnets based on the
pyrochlore structure. \cite{Zhitomirsky2003} They are particularly
suited for magnetic cooling as they exhibit the highest degree of
frustration and thus the highest cooling rates among all types of
geometrically frustrated magnets.  The degree of frustration is
directly connected to the number of soft magnetic excitations below
the saturation field. \cite{Zhitomirsky2003} A prototype material of
this class is Gd$_2$Ti$_2$O$_7$ where the largest magnetocaloric
effect is observed between 12\,T and 6\,T with a ${\rm d}T/{\rm d}B
\approx 0.5$~K/T. \cite{Sosin2005} This field range corresponds to a
crossover regime between a saturated and the disordered cooperative
paramagnetic state. \cite{Sosin2005} It was shown that the cooling
rate of Gd$_2$Ti$_2$O$_7$ even exceeds that of Gd$_3$Ga$_5$O$_{12}$
around the saturation field. However, deviations from the ideal
classical highly frustrated Heisenberg antiferromagnet limits the
lowest accessible temperature $T_{\rm min}$ during a cooling
process. One such correction is given by quantum fluctuations
(compare section \ref{sec:kag}). Indeed, for Gd$_2$Ti$_2$O$_7$ and
Gd$_3$Ga$_5$O$_{12}$ good agreement with a classical Heisenberg
model was observed for $B>B_s$, but deviations were found at low
temperatures for $B<B_s$. \cite{Sosin2005} For such reasons,
application of these systems is favored for temperatures down to
approximately $0.5$~K. \cite{Hepburn1995}

\subsubsection{Quantum-critical systems}

\label{sec:Qcrit}

Here we summarize recent results of the MCE on materials close to a
quantum critical point. This includes low-dimensional insulating
quantum magnets and metallic heavy-fermion materials.\\

\noindent
(i) Low-dimensional quantum magnets\\

\noindent There are a number of excellent realizations of quasi-1D
quantum magnets such as Cu(C$_4$H$_4$N$_2$)(NO$_3$)$_2$
\cite{Hammar1999} and KCuF$_3$ \cite{Lake2005} where it has been
demonstrated that quantum-critical Luttinger liquid (LL) behavior
governs the material's properties over wide ranges in temperature
and energy. For quasi-2D antiferromagnets the compounds
[Cu(pz)$_2$](ClO$_4$)$_2$, [Cu(pz)$_2$](BF$_6$)$_2$ and
[Cu(pz)$_2$(NO$_3$)](PF$_6$) are good realizations.
\cite{Woodward2007} However, in most cases the magnetic exchange
couplings are too large for a technical application and in the case
of KCuF$_3$ they are of order 100\,K, requiring magnetic fields of
order 100\,T to drive the system quantum critical.

A proof-of-principle experiment, demonstrating an increased cooling
rate around $B_s$, was performed by using a copper-containing
coordination polymer
[Cu($\mu$-C$_2$O$_4$)(4-aminopyridine)$_2$(H$_2$O)]$_n$ (labeled CuP
in the following) \cite{Wolf2011}, first synthesized by Castillo
{\it et al.}. \cite{Castillo2000} A comparison of the results of
magnetic susceptibility and specific heat measurements on single
crystalline material with model calculations \cite{Kluemper1998}
showed that CuP is indeed a very good realization of a uniform
spin-1/2 AFHC with an intra-chain coupling $J/k_B = (3.2\pm0.1)$~K.
\cite{Prokofiev2007} Compared to other excellent model substances of
this kind, such as copper pyrazin dinitrate, \cite{Hammar1999} CuP
excels by its comparatively small saturation field $B_s$ = 4.09\,T
(for $B \parallel b$) \footnote{For a new spin-1/2 chain substance
with an even smaller saturation field that has also been investigated with
regard to magnetocaloric properties, see Ref.~\cite{Tarasenko2014}.
}. This enables to study the MCE in the relevant
field range below about twice the saturation field by using standard
laboratory magnets.

\begin{figure}[bt]
%\centerline{\psfig{file=Fig2_02062014.eps,width=3.65in}}
\centerline{\includegraphics[width=4.65in]{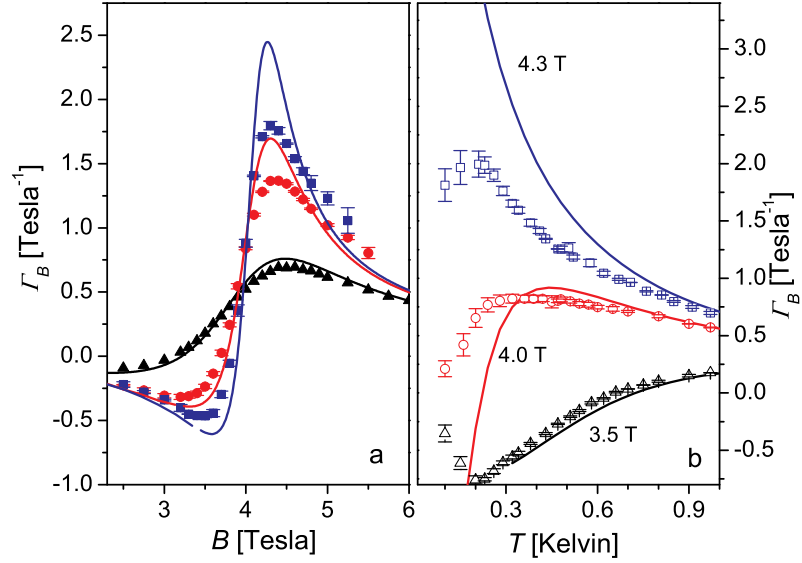}}
%\vspace*{8pt}
\caption{(a) Experimental data of the magnetic Gr\"{u}neisen parameter $\Gamma_B$ (symbols) of single crystalline
CuP (a copper-containing coordination polymer -- a good realization of a spin-1/2 antiferromagnetic Heisenberg chain) as a function of magnetic field ($B \parallel b$-axis) at $T$ = 0.97\,K (black full triangles) $T$ = 0.42\,K
(red full circles) and 0.32\,K (blue full squares).
The solid lines are the results of quantum Monte Carlo (QMC) simulations and exact diagonalization of a finite-size lattice of the spin-1/2 antiferromagnetic Heisenberg chain
for the corresponding temperatures (same colour code) by using $J$/$k_B$ = 3.2\,K. A $g$-value of 2.28
has been used in the calculations to account for a small misalignment of the crystals. (b) The magnetic Gr\"{u}neisen parameter
$\Gamma_B$ (symbols) of CuP as a function of temperature at constant fields below ($B$ = 3.5\,T and 4.0\,T) and above (4.3\,T)
$B_s$. The solid lines (same colour code as used for the experimental data) represent the results of model calculations
for the ideal AFHC for the corresponding temperatures. The figure is taken from \protect\cite{Wolf2011}.}
\label{GammaHeisenberg}
\end{figure}

In Fig.~\ref{GammaHeisenberg}a we show the MCE, approximated by
$\Gamma_B = T^{-1} \, (\Delta T/\Delta B)_{S \approx {\rm const}}$,
as a function of magnetic field for various constant temperatures.
The values of $\Gamma_B(B, T)$ are negative for $B < B_s$, implying
that here the material is cooling during the magnetization process,
in contrast to the behaviour of a simple paramagnet. Upon increasing
the field, $\Gamma_B(B, T)$ passes through a weak minimum for $T <
0.97$~K, changes sign, and adopts a pronounced maximum above 4~T.
The sign change in $\Gamma_B \propto$ ($\partial S$/$\partial
B$)$_T$ around $B_s$ implies the presence of a distinct maximum in
the magnetic entropy. This is expected around the QCP
\cite{Garst2005} and reflects the accumulation of entropy due to the
competition between the neighbouring ground states. The solid line
represents calculations of a parameter-free model. The theoretical
results are based on exact diagonalization and quantum Monte Carlo
simulations, performed by using the experimentally determined
exchange-coupling constant $J$/$k_B$ = 3.2\,K. \cite{Wolf2011}
Figure \ref{GammaHeisenberg}b exhibits the magnetic Gr\"{u}neisen
parameter as a function of temperature at constant field
$\Gamma_B$($T$, $B$ = const). The data were taken at fields below
($B$ = 3.5\,T and 4\,T) and above ($B$ = 4.3\,T) the saturation
field $B_s$. While the experimental results nicely agree with the
model calculations for the ideal system (solid lines) at higher
temperatures $T \geq$ 0.8\,K, the data progressively deviate from
the calculations with lowering the temperature. An anomaly is
clearly visible around 0.22\,K in Fig.~\ref{GammaHeisenberg}b at
3.5\,T, 4.0\,T, and 4.3\,T which is not expected from the
theoretical calculations. This feature is likely to be related to
the opening of a field-induced gap in the excitation spectrum
consistent with the presence of a weak Dzyaloshinskii-Moriya
interaction, allowed by symmetry in CuP, which gives rise to a
field-induced gap in the excitation spectrum.

Figure \ref{GammaHeisenberg} demonstrates that the magnetothermal
behaviour of CuP deviates from that of an ideal spin-1/2 AFHC especially at
low temperatures. This may limit the use of CuP for magnetic
cooling.

\begin{figure}[bt]
%\centerline{\psfig{file=Fig3_02062014.eps,width=3.65in}}
\centerline{\includegraphics[width=4.65in]{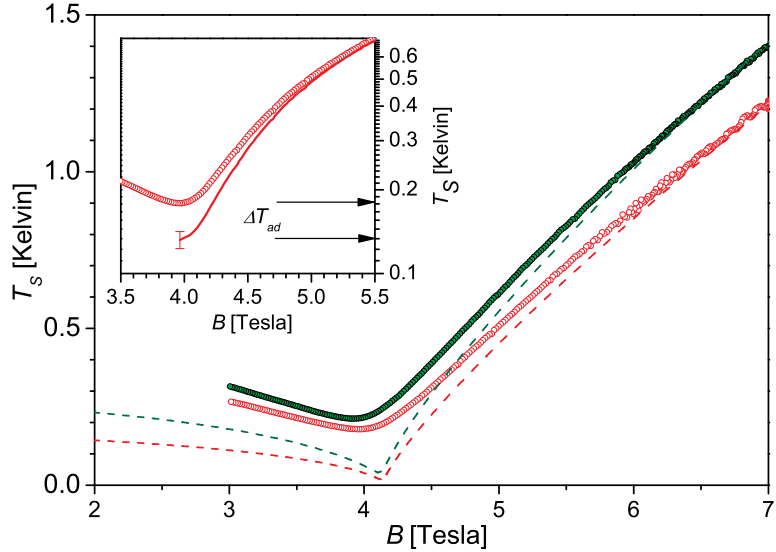}}
%\vspace*{8pt}
\caption{Sample temperature $T_s$ (symbols) measured upon demagnetizing a collection of CuP single crystals ($B \parallel b$-axis)
of total mass of 25.23\,mg under near adiabatic conditions. The initial parameters were set to $B_i$ = 7\,T and $T_i$ = 1.40\,K
(green symbols) and 1.21\,K (red symbols). Broken lines represent ideal isentropes, derived from the exact result for
the entropy of the spin-1/2 antiferromagnetic Heisenberg chain
\protect\cite{Trippe2010}
for the same initial parameters as used in the experiment.
%Fig. taken from \protect\cite{Lang2013}.
Inset: Experimental data with $T_i$ = 1.21\,K (red open symbols) together with data corrected for the parasitic heat load (red solid line).}
\label{CoolingHeisenberg}
\end{figure}

Therefore in Ref.\,\cite{Wolf2011} demagnetization experiments under
improved, near adiabatic conditions were performed to explore the
material's potential as a coolant. For these measurements, an array
of oriented single crystals of total mass of 25.23\,mg was used.
Starting from initial parameters $B_i$ and $T_i$, the field was
ramped down at a rate $\Delta B$/$\Delta t$ = -0.3\,T/min for $B
\geq$ 6\,T and -0.5\,T/min for $B <$ 6\,T. As
Fig.~\ref{CoolingHeisenberg} indicates, $T_s$ initially decreases
linearly with $B$ such as seen for simple paramagnets where $T$/$B$
= const. With further decreasing temperature, however, the cooling
process becomes superlinear in accordance with the model
calculations for the ideal system (broken line in
Fig.~\ref{CoolingHeisenberg}). This enhanced cooling rate is a
direct manifestation of quantum criticality. In this test
experiment, with a given weak thermal coupling to the surrounding
bath (sample holder) which is kept at a temperature of 1.2\,K, a
minimum temperature of 0.179\,K was reached for a starting
temperature $T_i$ = 1.21\,K. The effect of the finite parasitic heat
load due to non-adiabatic conditions, present to some extent in any
cooling experiment, has been estimated in Ref.~\cite{Lang2013} based
on simultaneous measurements of the thermal coupling of CuP to the
bath. According to these calculations a minimum temperature of
0.132\,K is expected under hypothetically ideal adiabatic conditions
which corresponds to a temperature reduction of $\Delta T_{ad} \sim$
50\,mK, see inset to Fig.~\ref{CoolingHeisenberg}. These experimental findings indicate that the main source for
the deviations between the cooling curve of CuP and the theoretical
calculations for $T_s$($B$), indicating an accessible minimum
temperature below 20\,mK, is likely the above-mentioned perturbing
interactions, especially the field-induced excitation gap due to the
Dzyaloshinskii-Moriya interaction. On the other hand, the good
agreement with the theory curve at higher temperatures, where these
interactions are irrelevant, shows that for those realizations of a
spin-1/2 antiferromagnetic Heisenberg chain where the perturbing
interactions are less strongly pronounced, cooling to much lower
temperatures should be possible. As promising candidates we
therefore suggest to focus on antiferromagnetic spin-1/2 chain
compounds with (i) a sufficiently weak intra-chain coupling $J$, and
(ii) magnetic centres related by inversion symmetry so that the
Dzyaloshinskii-Moriya interaction is not active.

\begin{figure}[bt]
%\centerline{\psfig{file=Fig4_02062014.eps,width=3.65in}}
\centerline{\includegraphics[width=4.65in]{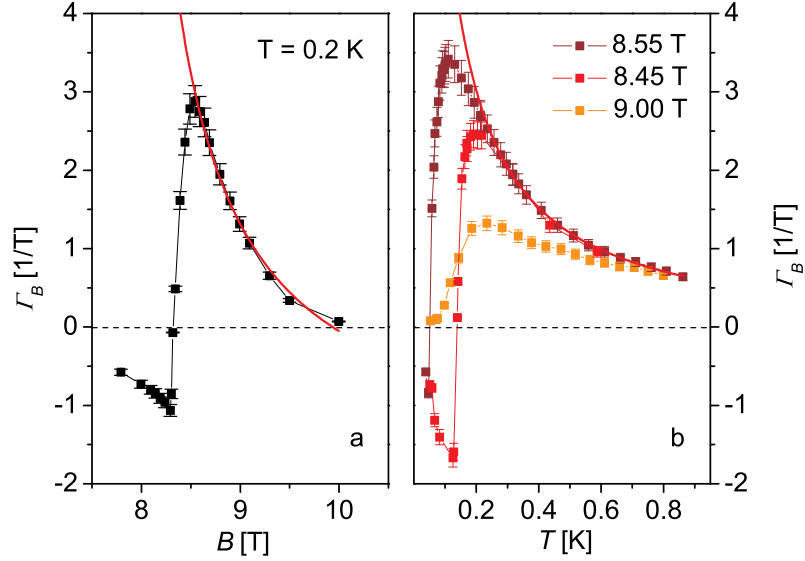}}
%\vspace*{8pt}
\caption{(a) Data for the magnetic Gr\"{u}neisen parameter $\Gamma_B$ of single crystalline Cs$_2$CuCl$_4$ for
$B \parallel a$-axis as a function of field at $T$ = 0.2\,K. (b) $\Gamma_B$ as a function of $T$ for
constant magnetic fields around the QCP. The solid red lines in both panels indicate an 1/$T$ dependence.
Figure taken from \protect\cite{Lang2013}.}
\label{GammaCsCuCl}
\end{figure}

As another example for a quantum-critical system, we mention triangular-type
frustrated quasi-2D Heisenberg antiferromagnets. Such systems are
expected to exhibit an even larger cooling capability as compared to
the AFHC as a consequence of the peculiar entropy landscape which is
a result of the geometric frustration of the material, compare section
\ref{sec:kag}. A good
realization of such a system is found in the layered
triangular-lattice Heisenberg antiferromagnet Cs$_2$CuCl$_4$ where
the frustration results from a dominant antiferromagnetic exchange
coupling constant $J/k_B = 4.3$~K \cite{Coldea1996} along the
in-plane $b$-axis and a second antiferromagnetic in-plane coupling
$J' \sim J$/3 along a diagonal bond in the $bc$-plane.
\cite{Coldea2002} Further couplings in this material include a weak
inter-plane interaction $J'' \sim J$/20 and a small anisotropic
Dzyaloshinskii-Moriya interaction $D \sim J$/20. \cite{Coldea2001}
This material has attracted considerable interest due to the 2D
character of its magnetic excitations in certain regions of the
$B-T$ phase diagram and its quantum-critical properties.
\cite{Coldea1996,Coldea2002,Coldea2001} The system exhibits a
field-induced quantum phase transition at a critical field $B_s$ of
about 8.5-9\,T, depending on the orienation of the magnetic field.
Here $B_s$ separates long-range antiferromagnetic order from a
fully-polarized ferromagnetic state.
%The QCP is located around 8.5\,T for the orientation $B \parallel a$ \cite{Garst2005,Coldea2001}.

Figures \ref{GammaCsCuCl}a,b exhibit the results of $\Gamma_B$ on
single crystalline Cs$_2$CuCl$_4$ as a function of magnetic field
(Fig.~\ref{GammaCsCuCl}a) at $T$ = 0.2\,K and temperature
(Fig.~\ref{GammaCsCuCl}b) at various constant fields around the QCP
which is located at around 8.5\,T for $B \parallel a$
\cite{Coldea2001} used in these experiments. The $\Gamma_B$ values
taken at 0.2\,K reveal a sign change at 8.32\,T, reflecting the
$B$-induced transition, in accordance with literature results on the
$B-T$ phase diagram. \cite{Radu2005} For $B < B_s$ $| \Gamma_B|$
is relatively small, indicating a flat entropy landscape in
the material's antiferromagnetically ordered state. However, upon
increasing the field to above $B_s$, $\Gamma_B$ changes sign and
adopts a pronounced maximum at about 8.53\,T. The sign change of
$\Gamma_B$ together with its critical enhancement just above $B_s$
and the approximate 1/$B$ reduction with increasing distance from
$B_s$ are clear signatures of quantum-critical behaviour.
\cite{Garst2005,Zhu2003} Figure \ref{GammaCsCuCl}b exhibits
$\Gamma_B$ as a function of temperature at constant magnetic fields
of $B$ = 8.45\,T and 8.55\,T and at $B$ = 9\,T. For the data taken
at 8.45\,T and 8.55\,T we find the same 1/$T$ dependence for 0.25\,K
$\leq T \leq$ 0.9\,K and a sign change in $\Gamma_B$ indicating the
proximity of a QCP. Similar to the observations for the AFHC
compound CuP, the deviations of $\Gamma_B$ from the ideal behaviour,
as revealed in the experimental data at low temperatures, are
assigned to the opening of a gap in the magnetic excitation
spectrum. The gap has a size of $\Delta$/$k_B$ $\sim$ 0.23\,K as
determined by inelastic neutron-scattering experiments.
\cite{Coldea1996} At $B$ = 9\,T, i.e., above the QCP, the
temperature dependence of $\Gamma_B$ is qualitatively different from
the behaviour observed around the QCP. There is no sign change in
$\Gamma_B$ and the experimental data exhibit a 1/$T$ dependence only
for temperatures above 0.7\,K as shown in Fig. \ref{GammaCsCuCl}b.
The gap in the magnetic excitation spectrum is clearly visible through the strong reduction of $\Gamma_B$ below 0.2\,K.\\

Apart from the effects of a small excitation gap, the observed
strongly enhanced $\Gamma_B$ values of Cs$_2$CuCl$_4$, indicating
large absolute variations of entropy around the QCP, suggest an
improved cooling performance of this frustrated quantum-spin system.

\begin{figure}[bt]
%\centerline{\psfig{file=Fig5_02062014.eps,width=3.65in}}
\centerline{\includegraphics[width=4.65in]{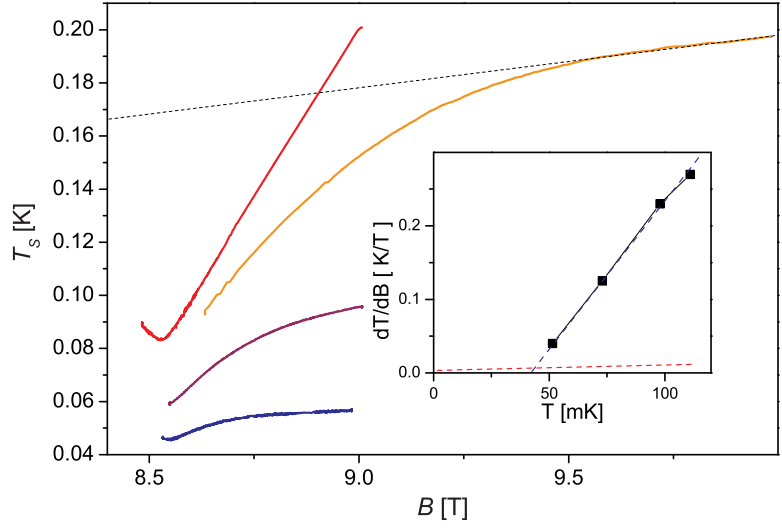}}
%\vspace*{8pt}
\caption{Experimentally determined cooling curves for single crystalline
Cs$_2$CuCl$_4$ obtained by demagnetizing the samples ($B \parallel a$-axis)
under almost adiabatic conditions. The starting temperature was $T_i$ = 0.2\,K while
cooling was performed from an initial field $B_i$ of 9\,T (red curve) and 10\,T (orange curve).
For comparison lowest-temperature cooling curves (violet and blue curves) are shown for $B_i$ = 9\,T and $T_i <$ 0.1\,K.
The dotted black line corresponds to the cooling curves of a paramagnet with the same initial parameters.
Inset: cooling rate d$T$/d$B$ of Cs$_2$CuCl$_4$ at a field of 8.65\,T at various temperatures (black squares),
as derived from the slopes of the various cooling curves at this field level, compared to that of a paramagnet (red circles)
indicating the enhanced cooling performance in the quantum-critical regime.}
\label{CoolingCurvesCsCuCl}
\end{figure}

Figure \ref{CoolingCurvesCsCuCl} shows a selection of cooling curves
for Cs$_2$CuCl$_4$ in a limited $B-T$ parameter range and compares
the cooling characteristics with that of a paramagnet. The four
cooling curves for Cs$_2$CuCl$_4$ are obtained with different
combinations of starting parameters $T_i$ and $B_i$ all of which are
located close to the QCP. As can be seen in the main panel of
Fig.~\ref{CoolingCurvesCsCuCl} a final temperature $T_f$ = 0.083\,K
is reached with a starting temperature $T_i$ = 0.2\,K and an initial
field of $B_i$ = 9\,T. According to Fig.~\ref{GammaCsCuCl}a, this
combination of parameter corresponds to a magnetic Gr\"{u}neisen parameter
$\Gamma_B \approx$ 1 T$^{-1}$ for Cs$_2$CuCl$_4$. The average
cooling rate d$T$/d$B$ of this curve is approximately 0.25\,K/T. For
comparison, a paramagnet demagnetized from the same initial
parameters would only reach a final temperature of $T_f$ = 0.19\,K,
corresponding to a cooling rate d$T$/d$B$ = 0.02\,K/T. Even with
$T_i$ = 0.2\,K and $B_i$ = 0.5\,T, initial parameters much more
suitable for paramagnets, $T_f$ will stay slightly above 0.1\,K.
This is a further experimental proof of the enhanced cooling
performance of a frustrated quantum-critical system. However, a
slight variation of starting parameters to regions in the $B-T$
phase diagram, where $\Gamma_B$ is no longer critically enhanced,
the cooling performance of Cs$_2$CuCl$_4$ is significantly altered
and the cooling rate becomes that of a simple paramagnet. For
example, by keeping $T_i$ = 0.2\,K and increasing $B_i$ to fields
above 9.5\,T, the cooling curve of Cs$_2$CuCl$_4$ (orange curve in
Fig.~\ref{CoolingCurvesCsCuCl}) becomes identical to that expected
for a paramagnet (black broken line). Also a reduced cooling rate is
found by keeping $B_i$ at 9\,T while decreasing $T_i$ to 0.096\,K
and 0.057\,K. For the latter starting temperature, the high-field
cooling behaviour around 9\,T also matches that of a paramagnet.
These experimental findings demonstrate that an enhanced cooling
rate for Cs$_2$CuCl$_4$ is confined to its quantum-critical regime,
whereas outside this regime on the high-field side, a performance
identical to that of a paramagnet is revealed. Another way of
visualizing this crossover in the cooling behaviour is provided by
following the slopes of the various cooling curves at a constant
field of 8.6\,T, i.e., about 0.1\,T above the critical field $B_s$,
cf. inset of Fig.~\ref{CoolingCurvesCsCuCl}. At this field the
cooling curves are, to a good approximation, linear in $B$, albeit
with distinctly different slopes. As shown in the inset of
Fig.~\ref{CoolingCurvesCsCuCl}, the cooling rate d$T$/d$B$ of
Cs$_2$CuCl$_4$ at $B$ = 8.65\,T (filled black squares) decreases
linearly with temperature and crosses d$T$/d$B$ of the paramagnet
(filled red spheres) at around 40\,mK for this field value.\\

\noindent
(ii) Heavy-fermion intermetallics\\

\noindent Heavy-fermion systems are intermetallic compounds, mostly
based on Ce, Yb, or U, where the 4$f$ or 5$f$ elements form a dense
Kondo lattice. The physical properties of these materials are
determined by strong electronic correlations and their ground state
depends sensitively on the balance between the on-site Kondo
screening and inter-site magnetic exchange coupling. In many
heavy-fermion compounds quantum-critical behaviour has been observed
and intensively investigated using thermodynamic methods.
\cite{Gegenwart2008,Gegenwart2010} The MCE has been investigated
down to lowest temperatures in the prototypical heavy-fermion metal
YbRh$_2$Si$_2$. \cite{Tokiwa2009} This system exhibits a
field-tuned QCP which is related to the suppression of an
antiferromagnetic order at a small critical field of $B_c$ =
0.06\,T. From magnetization measurements and specific heat data the
magnetic Gr\"{u}neisen parameter $\Gamma_{B}$ was determined for
YbRh$_2$Si$_2$. As expected by theory $\Gamma_{B}$ strongly
increases around the QCP, a property which makes such heavy-fermion
materials potential candidates for low-temperature magnetic cooling.
Compared to the quantum-critical magnetic insulators, heavy-fermion
metals may have a higher thermal conductivity at low temperatures,
which can be of relevance for applications at temperatures
significantly below 0.1\,K.

\begin{figure}[bt]
%\centerline{\psfig{file=Fig6_02062014.eps,width=3.65in}}
%\centerline{\includegraphics[width=3.65in]{Fig6_02062014}}
\centerline{\includegraphics[width=3.65in]{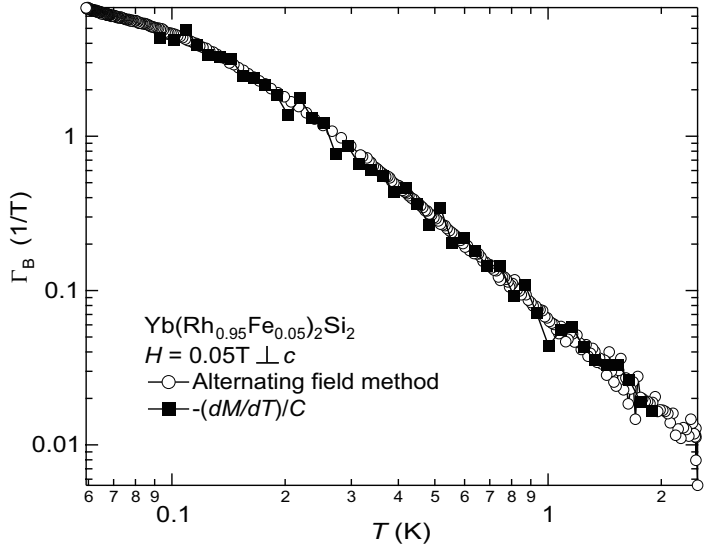}}
%\vspace*{8pt}
\caption{Magnetic Gr\"{u}neisen parameter
$\Gamma_{B}$ for the heavy-fermion material
Yb(Rh$_{0.95}$Fe$_{0.05}$)$_2$Si$_2$ as a function of temperature
data taken from Ref. \protect\cite{Tokiwa2011}. The solid squares are
calculated from magnetization and specific heat data whereas the
open circles are the results of a direct determination of the MCE
using an alternating field technique.}
\label{GammaHF}
\end{figure}

Figure \ref{GammaHF} shows the direct determination of the MCE for
the heavy-fermion material Yb(Rh$_{0.95}$Fe$_{0.05}$)$_2$Si$_2$
(open circles) using a newly developed method described in detail in
Ref.~\cite{Tokiwa2011}. The figure also displays the calculated
values of the MCE based on magnetization and specific heat data
(full squares). The experiment was performed above the QCP at the
critical field of 0.05\,T. As displayed in this figure $\Gamma_{B}$
tends to diverge upon approaching the QCP, reaching a value of about
6\,T$^{-1}$ at 60\,mK. Similar values of $\Gamma_{B}$ are also found in the
undoped material YbRh$_2$Si$_2$. \cite{Tokiwa2009}

\subsection{Cooling in ultracold gases}

%laser cooling: $\mu K$ regime (typically $S/N \approx 10 k_B$)   \\
%evaporative cooling : $S/N < \approx 3.6 k_B $ \\
For noninteracting Fermi gases, the lowest reported temperatures, achieved in $^6$Li samples
by sympathetic evaporative cooling with bosonic $^{23}$Na, are
$T = 0.05 \, T_F = 93 \,{\rm nK}$   \cite{Hadzibabic2003} , while for a noninteracting BEC of $^{23}$Na the
Picokelvin regime has been reached by pure evaporative cooling, with a final temperature of
$T \approx 500 \,{\rm pK}$. \cite{Leanhardt2003} \\
%Ketterle: below 500 pK ($T_c < 1 nK$), pure evaporative cooling of $^{23}$Na
%\cite{Leanhardt2003}. Higher relative temperature than in Ref. [30]
%of McKay/DeMarco. \\
For the weakly interacting regime (in a harmonic trap, before transfer into an optical lattice)
that is relevant for lattice experiments, $0.13T /T_F$ \cite{Joerdens2010} and $0.3T /T_c$ \cite{McKay2008} are the lowest reported \emph{relative} temperatures for ultracold fermions and bosons, respectively, which correspond to $S/N \approx 1k_B$ and $S/N \approx 0.1k_B$ for an ideal gas.\cite{McKay2011}\\
% In [30]: The average $T_c$ for the data in this paper is $0.13 \mu K$
%
%For strongly interacting Fermi gases, similar effective temperatures have been observed,
%and cooling down to to $S/N \approx 0.6k_B$ [113 Luo/Thomas] has been demonstrated.
%\emph{this seems to be colder than Joerdens below.
%different regimes: strongly vs. weakly interacting.}\\
%A possible method for removing the higher bands is to use Raman transitions to free particle states [147]. Simulations of non-interacting fermions suggest that temperatures as low as T /TF Å 0.001 can be attained if the proper potentials can be created [147]. \\
By using the spin-gradient demagnetization cooling technique in an optical lattice, isolated hyperfine spin distributions at
positive and negative effective spin temperatures of 50 pK have been prepared. In the same experiment, the spin subsystem has also
been used for cooling other (e.g., motional) degrees of freedom, leading to an equilibrated Mott insulator of $^{87}$Rb atoms at approximately 350 pK.
These temperatures are the lowest ones ever measured in any thermodynamic
system so far.\cite{Medley2011}

\section{Applications}

\label{sec:appl}

Nowadays magnetic refrigeration to temperatures below 100\,mK is a
well-established (even though not frequently used) cooling
technology for those applications where low costs, low complexity
and the ease of operation, combined with high reliability,  are in
demand.

However, an important field of application for magnetic cryocoolers
are space missions which, in turn, have driven an active development
of this technology. In the primary focus are astronomy missions
where low-temperature detectors and cooled telescopes are used.
These applications require coolers which ensure long hold times for
a given, fixed temperature below 100\,mK. Furthermore the coolers
can also support research in areas such as fundamental physics under
microgravity. There is an increasing demand for magnetic coolers in
space applications and therefore new developments have greatly
expand their capabilities and range of use. An important step are
multi-stage coolers with the ability to release heat at temperatures
that can be provided by mechanical cryocoolers used as precooling
stages. This makes it possible to operate the systems without the
complexity of using liquid cryogens that heretofore have limited the
mission lifetimes to 1-3 years.

\subsection{Solid state systems}

As described in the previous sections, the cooling performance of a
given material is determined by the low-energy sector of its
magnetic excitation spectrum, reflected in the entropy landscape as
a function of $B$ and $T$. For paramagnetic materials the residual
interactions of the magnetic ions lead to a reduced entropy and
eventually to long-range magnetic order at low temperatures.
Diluting the paramagnetic systems by the partial substitution of the
magnetic ions or by the incorporation of water into the crystal
structure can suppress $T_c$. However, these modifications also
reduce the density of magnetic ions, and by this, reduce the
materials' cooling capacity per unit volume which is the amount of
heat absorbed by the magnetic coolant ($\Delta Q_c$ in table
\ref{tab:1}). The crucial parameters, relevant in the selection of a
suitable paramagnetic refrigerant for a given application, are the
size of the spin $s$ of the magnetic centres, the magnetic ion
density and the material's ordering temperature $T_c$. The latter
quantity eventually determines the material's operating range. Thus
the goal is to use a material with $T_c$ close to, but below, the
desired operating temperature. For applications significantly below
100\,mK, materials necessarily have to have a low spin value and a
low ion density, which result in relatively low cooling capacity.

\begin{figure}[bt]
%\centerline{\psfig{file=fFig7_02062014.eps,width=3.65in}}
\centerline{\includegraphics[width=4.65in]{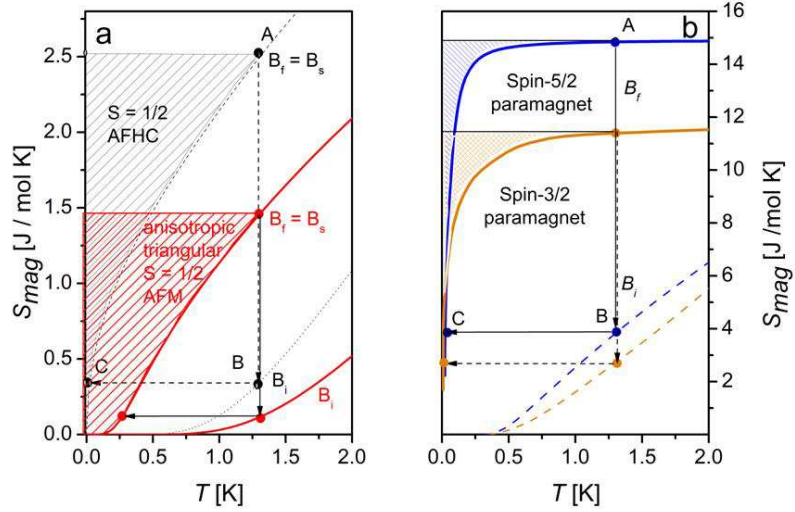}}
%\vspace*{8pt}
\caption{(a) Molar magnetic entropy $S_{\rm mag}$($T$, $B$ = const.) vs.\ $T$ calculated for the spin-1/2
antiferromagnetic Heisenberg chain ($s$ = 1/2 AFHC) with $J$/$k_B$ = 3.2\,K for an initial field
of $B_i$ = 7.14\,T (broken grey line) and a final field $B_f$ = $B_s$ = 4.09\,T (dotted grey line).
The path AB denotes an isothermal magnetization process at $T_i$ = 1.3\,K, a realistic temperature for
applications where precooling is provided by a pumped $^{4}$He bath, followed by an adiabatic demagnetization
(path BC). While the heat of magnetization, $\Delta Q_m$, corresponds to the area of the rectangle AB $\times$ BC,
the amount of heat that the material is able to absorb, $\Delta Q_c$, is given by the hatched area, cf. eq.\,\ref{DeltaQc}
below. For comparison, the figure includes $S_{\rm mag}$($T$, $B$ = const.) calculated for the frustrated
two-dimensional triangular lattice spin-1/2 Heisenberg antiferromagnet with parameters appropriate for
Cs$_2$CuCl$_4$. (b) $S_{\rm mag}$($T$, $B$ = const.) of the spin-5/2
paramagnetic salt Fe(NH$_4$)(SO$_4$)$_2\cdot$12H$_2$O
for $B_i$ = 2\,T (blue broken line) and $B_f$ = 0 (blue solid line) and the spin-3/2 paramagnetic salt
CrK$_4$(SO$_4$)$_2 \cdot$12H$_2$O for $B_i$ = 2\,T (orange broken line) and $B_f$ = 0 (orange solid line). The data are taken from
\protect\cite{Vilches1966,Pobell1992}. Panel (a) are taken from Ref.~\protect\cite{Lang2013}. }
\label{CoolingPerformance}
\end{figure}

The use of quantum-critical materials, as a new concept for magnetic
cooling, offers the possibility to get around some of the
limitations of paramagnets. Quantum-critical systems, which per se
lack long-range order, allow for large ion densities and thus
afford large cooling capacities down to (in principle) arbitrarily
low temperatures. In table \ref{tab1} the paramagnetic standard
materials for magnetic cooling below 100\,mK and their relevant
characteristics are listed together with the corresponding figures
of quantum-critical systems. The spin-3/2 (CPA) and 5/2 (FAA) are
the magnetic coolants for space applications
\cite{Shirron2007,Hagemann1999} whereas CMN with the lowest
transition temperature is only used for cooling in laboratories or
as a thermometer for low-temperature applications. An important
quantity, especially for space applications, is the heat of
magnetization $\Delta Q_m$ which has to be released to some
precooling stage. Another item, quantifying the cooling
performance, is $\Delta Q_c$, the heat the material is able to
absorb after adiabatic demagnetization, see also
Fig.~\ref{CoolingPerformance}. $\Delta Q_c$/$\Delta Q_m$ is the
efficiency factor. The data of $\Delta Q_m$ and $\Delta Q_c$ are
calculated from the magnetization and cooling process shown in
Fig.~\ref{CoolingPerformance}.

Table \ref{tab1} compiles material properties, including the spin
$s$, the ion density, and the magnetic ordering temperature $T_c$,
and some performance characteristics for the state-of-the-art
paramagnetic coolants ferric ammonium alum Fe(NH$_4$)(SO$_4$)$_2
\cdot$12H$_2$O (FAA), \cite{Cooke1949} chromic potassium alum
CrK(SO$_4$)$_2 \cdot$12H$_2$O (CPA), \cite{Bleaney1950} and cerium
magnesium nitrate (CMN). \cite{Cooke1953} These figures are compared
with those of quantum-critical systems, including the
antiferromagnetic spin-1/2 Heisenberg chain compound CuP and the
frustrated quasi-2D triangular-lattice antiferromagnet
Cs$_2$CuCl$_4$.

\begin{table}[tb]   %Table~1
\tbl{Material properties, including the spin $s$, the ion density,
and the magnetic ordering temperature $T_c$, and some performance
characteristics for the state-of-the-art paramagnets ferric ammonium
alum Fe(NH$_4$)(SO$_4$)$_2 \cdot$12H$_2$O (FAA)
%\cite{Cooke1949},
chromic potassium alum CrK(SO$_4$)$_2 \cdot$12H$_2$O (CPA)%
%\cite{Bleaney1950}
, and cerium magnesium nitrate
(CMN)
%\cite{Cooke1953}
in comparison with quantum-critical systems,
including the antiferromagnetic spin-1/2 Heisenberg chain compound
CuP and the frustrated quasi-2D triangular lattice antiferromagnet
Cs$_2$CuCl$_4$. $\Delta Q_m$ denotes the heat of magnetization which
has to be released to some precooling stage, and $\Delta Q_c$ the
heat the material is able to absorb after adiabatic demagnetization,
see also Fig.~\ref{CoolingPerformance}. $\Delta Q_c$/$\Delta Q_m$
is the efficiency factor.
\label{tab:1}}
{\begin{tabular}{@{}lccccc@{}} \Hline
\\[-1.8ex]
 {} & CMN & CPA & FAA & CuP & Cs$_2$CuCl$_4$ \\[0.8ex]
\hline \\[-1.8ex]
spin $s$ & 1/2 & 3/2 & 5/2 & 1/2 & 1/2 \\ ion density [cm$^{-3}$] & 1.05$\cdot$10$^{21}$ & 2.14$\cdot$10$^{21}$ & 2.15$\cdot$ 10$^{21}$ & 2.96$\cdot$10$^{21}$ & 4.32$\cdot$10$^{21}$ \\$T_c$ [K] & 0.002 & 0.009 & 0.026 & -$^{\rm a}$ & -$^{\rm a}$\\
$\Delta Q_m$ [mJ/cm$^{3}$] & {} & 37.0 & 50.7 & 14.5 & 8.4 \\
$\Delta Q_c$ [mJ/cm$^{3}$] & {} & 3.9 & 4.3 & 3.7 & 4.8\\
$\Delta Q_c$/$\Delta Q_m$ & {} & 0.11 & 0.09 & 0.26 & 0.57\\
%Stock picking schedule & Inventory control & {} \\
%Order priorities & Factory order control & Execution \\
%Scheduling & Machine (work-centre) control & {}\\
%Operation sequencing & Tool control$^{\rm a}$ & {} \\[0.8ex]
\Hline \\[-1.8ex]
\multicolumn{3}{@{}l}{$^{\rm a}$ used for cooling above $B_c$.}\\
\end{tabular}}
\label{tab1}
\end{table}

Table \ref{tab1} shows that for the paramagnetic materials FAA, CPA and CMN a decreasing ion density is accompanied by a significant reduction in $T_c$. At the same time, a comparison between FAA and CPA, sharing about the same ion density, clearly illustrates the influence of the spin value on $T_c$. Reducing $s$ = 5/2 (FAA) to $s$ = 3/2 (CPA), while keeping the ion density practically unaffected, leads to a decrease in $T_c$ from 26\,mK (FAA) down to 9\,mK (CPA). In contrast to the paramagnets, the quantum-critical systems CuP (1D AFHC) and Cs$_2$CuCl$_4$ (2D AFM) exhibit significantly higher ion densities; for Cs$_2$CuCl$_4$, the ion density is a factor of about two larger compared to that of the paramagnets. Although the quantum-critical materials are $s$ = 1/2 compounds, they exhibit a similar or even higher cooling capability $\Delta Q_c$ compared to the standard paramagnets. The main advantage provided by the quantum-critical systems is the amount of heat released to the high-temperature bath (precooling stage) $\Delta Q_m$. For the process shown in Fig.~\ref{CoolingPerformance}, which assumes a bath temperature of 1.3\,K, these values are significantly smaller for the quantum-critical materials compared to the paramagnets. As a consequence, the efficiency $\Delta Q_c$/$\Delta Q_m$ of the cooling process for the quantum-critical systems largely exceeds the corresponding values for the paramagnets. This aspect can be of crucial importance for space applications. Given a limited capacity of the higher-temperature bath in these space missions, the higher efficiency of the cooling process directly leads to a significantly increase in the life time of the experiment.

Representative parameters characterizing the performance of a magnetic cryocooler include (i) the operating
range, in particular the lowest temperatures $T_f$ which can be obtained, (ii) the ``hold time'' of the coolant, which is inversely proportional to its cooling power and (iii) the efficiency $\Delta Q_c$/$\Delta Q_m$. The latter quantity can be read off the entropy-temperature diagram.

Figure \ref{CoolingPerformance} compiles curves of the molar magnetic entropy
$S_{\rm mag}$($T$, $B$ = const.) vs.\ $T$ for the various materials discussed here and listed in Table \ref{tab1}. This includes calculations for the spin-1/2 antiferromagnetic Heisenberg chain with $J$/$k_B$ = 3.2\,K at an initial field $B_i$ = 7.14\,T (broken line) and a final field $B_f$ = $B_s$ = 4.09\,T (dotted line). The path AB denotes an isothermal magnetization process followed by an adiabatic demagnetization (path BC) down to the final temperature of approximately 24\,mK. The heat of magnetization, $\Delta Q_m$, produced by magnetizing the material at the initial temperature $T_i$, corresponds to the area of the rectangle AB $\times$ BC, whereas the heat $\Delta Q_c$ the material can absorb after demagnetization to the final field $B_f$
\begin{equation}
\Delta Q_c = \int_{T_{f}} ^{T_{i}} T \left(\frac{\partial S}{\partial T} \right )_{B_{f}} dT
\label{DeltaQc}
\end{equation}
is given by the hatched area. The efficiency $\Delta Q_c$/$\Delta
Q_m$ in the temperature range indicated amounts to 26\%\ for the
spin-1/2 antiferromagnetic Heisenberg chain. For comparison, we show
in Fig.~\ref{CoolingPerformance} $S_{\rm mag}$($T$, $B$ = const.)
calculated for the anisotropic version of the two-dimensional
triangular-lattice spin-1/2 Heisenberg antiferromagnet with
$J$/$k_B$ = 3$J'$/$k_B$ = 4.3\,K, at $B_i$ = 12.0\,T and $B_f$ =
$B_s$ = 8.55\,T, i.e., parameters appropriate for Cs$_2$CuCl$_4$\footnote{
A chain mean-field theory turned out to be more appropriate in this
case than a plain exact diagonalization as in section \ref{sec:kag}
for the isotropic triangular lattice.
see Ref.~\cite{Lang2013} for more details.}. For this
strongly frustrated system and the temperature range indicated an
efficiency $\Delta Q_c$/$\Delta Q_m$ as high as about 57\%\ is
obtained.

These numbers for the quantum-critical systems can be compared with
that of the spin-5/2 paramagnetic salt Fe(NH$_4$)(SO$_4$)$_2
\cdot$12H$_2$O (FAA). Figure \ref{CoolingPerformance} shows $S_{\rm
mag}$($T$, $B$ = const.) for $B_i$ = 2\,T (blue broken line) and
$B_f$ = 0 (blue solid line) corresponding to $T_i$ = 1.3\,K and
$T_f$ = 22\,mK. We find $\Delta Q_m$ = 50.71\,mJ/cm$^{3}$ and
$\Delta Q_c$ = 4.33\,mJ/cm$^{3}$ which corresponds to an efficiency
$\Delta Q_c$/$\Delta Q_m$ of only 9\%. In a similar procedure we
obtained for the spin-3/2 paramagnet CrK(SO$_4$)$2 \cdot$12H$_2$O
(CPA) $\Delta Q_m$ = 39.02\,mJ/cm$^{3}$ and  $\Delta Q_c$ =
3.89\,mJ/cm$^{3}$. This results in an efficiency factor $\Delta
Q_c$/$\Delta Q_m$ of about 11\%.

\section{Perspectives}

In this article we review recent developments in cooling
technologies for condensed matter and cold gas experiments. The
demand for new coolants in these areas arises from several factors
which are different for both fields. For condensed matter systems,
the reasons mainly lie in the restricted availability of Helium,
especially $^3$He, and the need to optimize the cooling performance
with regard to space, weight or the ease of operation. On the other
hand, for ultracold gases, it is the desire for reaching low enough
temperatures to eventually access solid state-type ordered states
such as quantum magnetism or $d$-wave superfluidity.

The new approach discussed here is based on taking advantage of
many-body interactions in the system to be cooled. We discuss some
proof-of-principle demonstrations for cooling materials through
quantum criticality and discuss the effects of geometric
frustration. By comparing experimental results with theoretical
calculations on model systems, we are able to identify limiting
factors for the real systems an indicate routes towards next
generation many-body coolants. We demonstrate that these systems
have the potential to outperform state-of-the-art conventional
refrigerant materials.

For ultracold gases we have discussed two recent approaches towards
adiabatic cooling of the many-body system, based on spin-gradient
demagnetization and the Pomeranchuk effect, which directly take into
account interparticle interactions in an optical lattice. Future
developments along these directions will likely involve reduced
spatial mass- and entropy-transport, in order to reduce
non-adiabatic effects which limit cooling efficiency, and local
addressing techniques such as the optical quantum gas microscope.

\section*{References}

%References in the text are to be numbered consecutively in Arabic
%numerals, in the order of first appearance. They are to be typed in
%superscripts after punctuation marks, e.g.

%(1) $ \qquad $ ``$\ldots$ in the statement.\cite{5}''

%(2) $ \qquad $ ``$\ldots$ have proven\cite{5} that this equation $\ldots$''

%\noindent
%This is done using the command: ``$\backslash$cite\{name\}''.

%\vspace*{6pt}
%\noindent
%When the reference forms part of the sentence if should not
%be superscripts, e.g.

%(1) $ \qquad $  ``One can deduce from Ref.~\refcite{5} that $\ldots$''

%(2) $ \qquad $  ``See Refs.~\refcite{1}--\refcite{3}, \refcite{5}
%and \refcite{7} for more details.''

%\vspace*{6pt}
%\noindent
%Typeset references in 9pt Times Roman.
%\vspace*{3pt}

\end{document}